\documentclass[runningheads,a4paper]{llncs}

\usepackage{amssymb}
\usepackage{graphicx}
\usepackage{slashbox}
\usepackage{multirow}

\usepackage{hyperref}
\usepackage{url}
\urldef{\mailsa}\path|Karydithalia@gmail.com|
\urldef{\mailsb}\path|kmarg@uom.gr|

\begin{document}

\mainmatter  

\title{Parallel and Distributed Collaborative Filtering: A Survey.}


%

\author{Efthalia Karydi \and Konstantinos G. Margaritis}
\authorrunning{Parallel and Distributed Collaborative Filtering: A Survey.}

\institute{ University of Macedonia, Department of Applied Informatics\\
Parallel and Distributed Processing Laboratory\\
156 Egnatia str., P.O. Box 1591, 54006 Thessaloniki, Greece
\mailsa\\
\mailsb\\
}

\maketitle

\begin{abstract}

Collaborative filtering is amongst the most preferred techniques when implementing recommender systems. Recently, great interest has turned towards parallel and distributed implementations of collaborative filtering algorithms. This work is a survey of the parallel and distributed collaborative filtering implementations, aiming not only to provide a comprehensive presentation of the field's development, but also to offer future research orientation by highlighting the issues that need to be further developed.

\end{abstract}

\section{Introduction}
\label{intro}

The quality of a recommender system's output is highly depended on the quantity of used data. The more data is available in a recommender system, the better will be the recommendation. Having to deal with continuously growing amounts of data, the design of parallel and distributed recommender systems has become necessary. The parallel and distributed computing techniques can be combined with each other to the purpose of exploiting their advantages and various modifications can be applied to the existing algorithms in order to fit better to the requirements of the used techniques. Furthermore, taking advantage of the heterogeneous infrastructures that are available is crucial for the development of high quality recommender systems. Thus, the study and design of parallel algorithms and implementations that will address the emerged problems and exploit the advantages of new technologies is important.

Among the benefits that are expected to be gained through the usage of parallel and distributed systems on the field of recommender systems are the following:
\begin{itemize}
\item Faster result delivery. The execution of online algorithms will be eased without efficiency loss.
\item Greater amounts of data can be used, fact that is expected to lead to greater efficiency.
\item Facilitate the simultaneous execution of different algorithms as long as the merging of their results. Therefore, the use of different data sources will be easier, as will be the variety of item types that can be recommended.
\item Privacy issues are better confronted on distributed systems. User trust on the recommender system is expected to increase.
\item Fault tolerant systems. If measures are taken to confront node failures, in case of such event, the overall system crush will be avoided.
\item Development of new and existing algorithms will be accomplished through the study for the choice, design and parallelization of the suitable algorithm.
\end{itemize}

Each parallel and distributed computing technique has advantages and disadvantages that must be considered in order to choose the most appropriate technology or an adequate combination to cope each problem.  Distributed implementations are adequate for privacy conserving that can augment a user's trust to the system but the communication cost among nodes may be high and even dominate the performance. Multithreading achieves fast runtimes but special care must be given to avoid memory conflicts and race conditions. The use of frameworks for massively parallel processing augments the processing speed and facilitates big data handling, yet the algorithm must be adequate for implementation over the selected framework  or must be appropriately modified. GPU usage can result to impressively fast processing as far as the algorithm employs matrix-vector computations. Memory accesses must be controlled to achieve the best performance possible. The selection of the appropriate architecture to be used depends on the problem that is faced and on the algorithm that is chosen for parallelization. The parallel and distributed computing techniques must be carefully chosen to help improve the overall performance.

\subsection{Basic Terminology}

Recommender systems are mechanisms that are used to produce item recommendations to their users. Their purpose is to make personalized recommendations that will be  interesting and simultaneously  useful for the users. This fact consists the big difference between recommender systems and information retrieval search engines \cite{RicciHand}.

Recommender systems can use a variety of data, such as the ratings that the users provide to the system for the system's items, user demographic information or product's content data. Data is exploited in the best possible way, in order to provide satisfactory recommendations or predictions. The output can be a list of item recommendations or a prediction of the value of the rating that a user would give to an item.

Recommender systems are especially useful to commercial applications due to the fact that they provide a means by which companies can effectively promote their products. Yet the interest in recommender systems is not centered only to commercial applications, but also to academic research that is still conducted keeping intense the researchers' interest  in their effort to face the challenges of improving the algorithms and the recommendation process and quality.

Recommender systems are categorized to the following classes, according to the techniques that are applied. \textbf{Collaborative Filtering} (CF) recommender systems exploit the fact that users with similar past preferences are likely to have common preferences again. \textbf{Content-based}  recommender systems calculate item similarities based on item's features. \textbf{Demographic} recommender systems use the users' demographic information. \textbf{Knowledge-based} recommender systems take advantage of specific domain knowledge that specifies to what extent the item is useful for the user. \textbf{Community-based} recommender systems provide recommendations based on the preferences of groups of users that have some common properties. All the above categories can be combined with each other and the recommender system that belongs to more than one categories is called \textbf{Hybrid} recommender system.
 
Collaborative filtering techniques are among the most popular techniques that are applied to recommender systems \cite{sarwar_e_com}. Collaborative filtering recommender systems are further classified into \textbf{model- based} and \textbf{memory-based}. \textbf{Hybrid} collaborative filtering recommender systems have been also developed, that combine model and memory-based methods. The difference of these categories is that memory-based algorithms use the entire dataset to make predictions, while model-based algorithms use a part of the data as a training set to create a model and then they use the model to create the predictions. In \cite{survey_cft} can be seen the collaborative filtering algorithms that belong to the above categories.

Although the field of recommender systems has been significantly developed, some problems still remain unsolved. Great concern is given to aspects such as the quality of the recommendations, the sparsity of the data, scalability, synonymy and how to cope with new users and items, which are issues that require attention since the beginning of the recommender systems' research \cite{sarwar_e_com}, \cite{pdp}.

The recommendations need to both attract the user's interest and be useful. The items that a user has already purchased should not be recommended again, as well as the items that are not according to the user's taste. By providing high quality recommendations, the user's trust to the recommender system is augmented and he is likely to continue using it.

The data sparsity is a growing problem that still needs to be faced. Usually the information that the users provide to the recommender system is very few considering the abundance of items that exist. This fact leads to very sparse data that delay the overall performance. Although many techniques have been developed to cope with data sparsity, it still remains a hot issue amongst the recommender system's problems.

Both the number of users and items are continuously growing. Thus, the need for fast and scalable computations is important. Nowadays, recommendations are expected to be produced extremely fast in order for a recommender system to be able to function properly online. Great effort must be given to develop efficient and scalable algorithms.

The difficulty to distinguish the latency among items that can have similar names but different content or completely different names but the same subject, is called the synonymy problem. The challenge of coping with the synonymy problem as long as the challenges to provide recommendations to users who are new to the system or who do not have a consistent taste similarity with any group of people, still require attention.

Other challenges that are concerning the recommender systems' research are the integration of methods to cope with long and short-term preference changes and the evaluation of recommender systems \cite{cha_rs_riedl}. Evaluating recommender systems under a common framework has been proved a hard task. Although some metrics are preferred to most of the existing approaches, questions still remain on how recommender systems should be evaluated.

The variety of technologies that exist can provide great advantages. To exploit them in an efficient fashion, the usage of heterogeneous systems has augmented. Thus, the algorithms should be redesigned to adjust well with the needs that emerge from the usage of heterogeneous systems.

Although research in the field of recommender systems is conducted over twenty years, the issues that still offer ground for improvement are not few. To cope with data abundance and to keep the time needed for the recommendations low, parallel and distributed systems are  more and more used.  In the following sections the approaches to recommender systems that employ parallel and/or distributed techniques will be surveyed in order to provide a concise view of the developments of the field and to highlight the factors that require further research.

\subsection{Collaborative Filtering Algorithms} 

Collaborative filtering algorithms are very popular among the existing approaches to recommender systems. The main idea of Collaborative filtering is that if a user has similar preferences with other users, then he will probably like items that other users with similar taste have liked and he is not aware of. A user's preference to an item is usually expressed with the rating that he gives to evaluate it. The collected ratings are used to calculate the similarity of the preferences of the users, and items are recommended based on the similarity value among two users.

Collaborative filtering techniques are classified to memory-based and model-based techniques \cite{survey_cft}. 

\subsubsection{Memory-based techniques} are also mentioned as neighbourhood-based methods. The entire dataset is used to calculate the similarity of the users with the active user. As active user is referred the user for whom the recommendation is produced. Then, a neighbourhood is formed by the k users that are most similar to the active user. Finally, the predictions of the ratings that the active user would give to the items are computed. The similarity is more often measured by Pearson Correlation Coefficient or by Cosine Vector Similarity \cite{pdp}, which are both variants of the inner product. The most popular algorithms that belong to this category are the item-based, the user-based and the Slope One algorithm. These algorithms can employ any of the similarity measures. The user and item-based algorithms are often encountered in top-N approaches, where a set of N items is recommended.



\subsubsection{Model-based techniques} use a part of the dataset to train a model and they produce the predictions according to the model. The objective of the model is to represent the user's  behaviour recognizing the behaviour patterns that occur on the training set and benefit from the observed patterns to create predictions for the real dataset. Various Machine Learning and Data Mining algorithms are used to create the model.

Linear algebra methods, such as Singular Value Decomposition (SVD), Principal Component Analysis (PCA), Latent Semantic Analysis (LSA), Latent Dirichlet Analysis (LDA), Stochastic Gradient Descent (SGD) and Alternating Least Squares (ALS) are very often used to represent users and items by means of an $f$-dimensional latent factor space. Models based on matrix factorization techniques are often preferred because they offer high accuracy and scalability \cite{RicciHand}. Other model-based techniques are Bayesian Networks, Clustering methods and Association Rule-based methods \cite{survey_cft}.

In table \ref{table:model} there is a list of the algorithms that have been implemented using parallel and distributed techniques, as long as the acronyms that will be used to the rest of this article.

\begin{table}[htp]
\scriptsize
\begin{center}
\begin {tabular}{|c|c|}
\hline
Algorithm & Description \\
\hline
SVD & Singular Value Decomposition \\
\hline
PCA & Principal Component Analysis \\
\hline
LSA & Latent Semantic Analysis \\
\hline
LDA & Latent Dirichlet Analysis \\
\hline
PLDA & Parallel Latent Dirichlet Analysis\\
\hline
SGD & Stochastic Gradient Descent \\
\hline
PSGD & Parallel Stochastic Gradient Descent \\
\hline
ASGD & Asynchronous Stochastic Gradient Descent\\
\hline
DSGD++ & Distributed Stochastic Gradient Descent ++\\
\hline
DSGD-MR & Distributed Stochastic Gradient Descent-MapReduce \\
\hline
FPSGD & Fast Parallel Stochastic Gradient Descent \\
\hline
ALS & Alternating Least Squares \\
\hline
ALS-WR & Alternating Least Squares with Weighted Regularization \\
\hline
PALS & Parallel Alternating Least Squares  \\
\hline
DALS & Distributed Alternating Least Squares \\
\hline
Wals& Weighted Alternating Least Squares\\
\hline
CCD++ & Coordinate Descent \\
\hline
CWSO & Clustering Weighted Slope One \\
\hline
NNMF & Non-negative Matrix Factorization\\
\hline
CAPSSR & Context aware p2p service selection and recovery\\
\hline
PLSI & Probabilistic Latent Semantic Indexing \\
\hline
BPTF & Bayesian Probabilistic Tensor Factorization\\
\hline
MFITR & Matrix Factorization item taxonomy regularization\\
\hline
RBM-CF & Restricted Boltzmann Machine- Collaborative Filtering \\
\hline
\end{tabular}
\caption{Acronyms}
\label{table:model}
\end{center}
\end{table}
 
\subsection{Evaluation Metrics}

How to evaluate  recommender systems is an issue that attracts great interest. Recommender systems can have various properties, such as being trustful, recommend novel, useful and interesting products, and being scalable. When designing a recommender system, one should decide which of the factors that characterize the recommender system are important for his implementation, and therefore, should select the adequate evaluation metrics to test whether the implementation meets the required criteria. A great variety of measures exists to evaluate each of the properties that a recommender system can have.  The difficulty of applying a common evaluation framework for all the recommender systems is revealed by considering the polymorphic nature that a recommender system can have and the variety of metrics.

One of the most important evaluation measurements is accuracy. Accuracy can measure how well a recommender system predicts a rating and is measured by means of Mean Absolute Error (MAE) or Round Mean Squared Error (RMSE). Measures also exist, that express how often a recommender system makes good or wrong recommendations. Metrics that classify accuracy are the F-measure, precision, recall, Receiver Operating Characteristic (ROC curves) and Area Under ROC Curve (AUC)\cite{metrics}.

Since the fast delivery of results is very important, time is an aspect that is often measured. Usually the total elapsed time is measured and the time of various tasks, such as the prediction delivery, the computation and the communication time is analysed. Furthermore, when parallel and distributed programming techniques are used, the corresponding metrics, such as speedup and isoefficiency  are also employed.

\subsection{Datasets}
\label{datasets}

In this section will be briefly presented the characteristics of the datasets that are used in the majority of the approaches discussed in the present work. A great variety of datasets is used in recommender systems' research. Some of them contain demographic data or timestamps and other emphasize in associations amongst the users. Also, different order of magnitude and diversity on the rating scale, as long as variety in sparsity and attributes provided in each dataset consist reasons for which the use of more than one datasets to evaluate a recommender system is fundamental.

One of the most commonly used datasets is the Netflix dataset, which was used for the Netflix Prize competition \cite{Netflix_Prize}. The dataset contains over 480000 users 17000 items and 100 million ratings. Unfortunately, the Netflix dataset is no longer available as is the EachMovie dataset.

GroupLens Research \cite{Movielensds} have released the MovieLens datasets, which are offered in various sizes shown in table \ref{table:datasets}. The MovieLens 10M dataset has been recently extended to MovieLens 2k, which associates the movies of MovieLens dataset with their corresponding web pages at Internet Movie Database (IMDb) \cite{imdb} and Rotten Tomatoes movie review system \cite{RottenTomatoes}. This dataset as long as the Delicious 2k and Last.fm 2k datasets, obtained from \cite{delicious} and  \cite{lastfm}, were released at the  2nd International Workshop on Information Heterogeneity and Fusion in Recommender Systems (HetRec 2011) \cite{Cantador:RecSys2011}.

The Book-Crossing dataset \cite{bookcross} contains ratings about books from 278858 users and demographic information.  Content information is also provided. A dense dataset is the Jester dataset, which contains data from the Jester Online Joke Recommender System \cite{jester}. Unfortunately this dataset contains only 100 items. The Yahoo! Music dataset \cite{yahoods} is also amongst the most used datasets. It was used for the KDD-Cup '11 contest. The ratings in this dataset are given to four different types of items (tracks, albums,artist, genres) and date and timestamp are provided in the track 1 dataset. The Flixster dataset \cite{flixster} contains ratings on movies and links amongst the users. In table \ref{table:datasets} can be seen the statistical information of the most commonly used datasets.

Timestamps are provided in the majority of the above datasets, except MovieLens 2k, Book-Crossing, Jester and EachMovie datasets. Demographic information is given in MovieLens, Book-Crossing and EachMovie datasets. To the last two datasets content information is also available and MovieLens 2k contains movie information. Delicious 2k and Last.fm 2k datasets provide social networking amongst the users. Depending on what is the main consideration of each experiment, a different dataset may be the most adequate. 

The main conclusion that results from the variety of the datasets that are used is that to be sure that an evaluation of a recommender system is accurate or that a comparison between various recommender systems is fair enough, more than one datasets have to be used.

\begin{table}[htp]
\scriptsize
\begin{center}
\begin {tabular}{|c|c|c|c|c|c|}
\hline
Dataset&Users &Items  &Ratings &Scale & Value \\
\hline
Netflix& 480,189  &17,770  &100,000,000 & 1-5 & integer\\
\hline
MovieLens 100k&943  &1,682  &100,000 & &  \\
MovieLens 1M &6,040  &3,900  &1,000,000 &1-5 &integer \\
MovieLens 10M&71,567  &10,681  &10,000,000 & &  \\
\hline
MovieLens 2k&2,113  &10,197  &855,598 &0-5 & real \\
(HetRec 2011)&  &  & & &  \\
\hline
Book-Crossing&278,858  &271,379  &1,149,780 &1-10 &integer \\
&  &  & & &  \\
\hline
Jester& 73,496 &100  & 4,100,000 & (-10) - (+10)& real \\
\hline
EachMovie &72,916  &  1,628 & 2,811,983 &0-5 & integer \\
\hline
Yahoo! music&  &  & & &  \\
KDD Cup 2011&  &  & & &  \\
track 1& 1,000,990 &624,961  &262,810,175 &1-5 &integer \\
track 2&  249,012 & 296,111  & 61,944,406 &1-5 &integer  \\
\hline
Flixster& 2,523,386 &49,000  & 8,200,000 & 1-5& real \\
\hline
Delicious 2k & 1,867 & 69,226  & \_ & \_& \_ \\
(HetRec 2011)&  & URLs & & & \\
\hline
Last.fm 2k& 1,892 & 17,632&\_ &\_ &\_  \\
(HetRec 2011)&  &artists  & & &  \\
\hline
\end{tabular}
\caption{Basic Information on Datasets}
\label{table:datasets}
\end{center}
\end{table}

\subsection{Classification Scheme}



The rest of this article is organized as follows. In section \ref{related} is provided a brief collection of the survey approaches found in literature, that concern recommender systems. As can be noticed, none of these works deals with parallel and distributed collaborative filtering recommender systems. In section \ref{surveydistributed} are presented the distributed implementations. Section \ref{parallel} concerns the parallel implementations separating them in three categories, according to whether they are implemented on distributed memory environments, on shared memory environments or whether they take advantage of GPU accelerators. Platform-based approaches are discussed in section \ref{surveyframeworks} and in section \ref{heterogeneous} are presented the heterogeneous approaches that belong to more than one of the above categories. In all sections, the implementations are classified according to which type of collaborative filtering belongs the algorithm that is implemented. The structure according to which the implementations are classified can be seen in table \ref{table:scheeme}. In the same table can also be seen the number of implementations that have been classified to each category. Finally, in section \ref{conclusions} the conclusions of the survey are presented.

To the best of our knowledge, the present work consists the first attempt to collect the parallel and distributed implementations of collaborative filtering recommender systems. Studying the existing implementations is expected to lead to the indication of further study sectors and to highlight the trends of the recent research, as long as the gaps and the difficulties of the field.

\begin{table}[htp]
\scriptsize
\begin{center}
\begin{tabular}{|c||c|c|c|}
\hline
 & \multicolumn{3}{c|}{\textbf{Collaborative Filtering}}\\
\cline{2-4}
 & \textbf{Memory-based} & \textbf{Model-based} & \textbf{Hybrid}\\
\hline\hline
	&	&	&	\\
\textbf{Distributed}				&13	& 4	& 6	\\
				&	&	&  	\\
\hline\hline
\textbf{Parallel}&	&	&	\\
	&	&	&  	\\
Distributed Memory	& 	& 7 & 1	\\
		&	&	&  	\\	
		&	&	&  	\\			
Shared Memory	& 1	& 6	&	\\
	&	&	&  	\\
			&	&	&  	\\	
		
GPU		&	4 & 	 9 &  	\\
	&	&	&  	\\	
\hline\hline
&	&	&	\\
\textbf{Platform-based}	&	7 & 10	& 1	\\
	&	& &  	\\
\hline\hline
&	&	&	\\
\textbf{Heterogeneous}	& 2 & 2	&	\\
	&	&	&  	\\
\hline
\end{tabular}
\caption{Classification of the Implementations}
\label{table:scheeme}
\end{center}
\end{table}


\section{Related Work}
\label{related}


This section is devoted to briefly outline the surveys concerning recommender systems. Recommender systems that combine different recommendation techniques are presented in one of the first surveys \cite{Burke2002}. A comparison among the different recommendation techniques is provided and their advantages and disadvantages are discussed. Also, the different hybridization methods are described. The existing hybrid approaches are briefly presented and a hybrid recommender system that combines knowledge-based recommendation and collaborative filtering is introduced. Experiments are conducted on the proposed recommender system using data derived from the web server's log. This survey proved that there were many combinations of techniques to be explored and outlined the needs of the field of hybrid recommender systems.

One of the early surveys addressing recommender systems is \cite{Adomavicius2005}. Recommender systems are classified into three categories. Content-based, collaborative and hybrid implementations. The constraints of each category are discussed and possible ways to improve the recommendation methods are proposed.

In \cite{ClassRSsurvey} the application domain of recommender systems is analysed. Almost 100 recommender systems are classified and the majority of them belong to the web recommendation, movie/TV recommendation and information/document recommendation application domains. Recommender systems are classified into six categories according to their functionality. The advantages and disadvantages of four of those categories are discussed.

A survey of the work in the field of web recommender systems is accomplished in \cite{Kumar10collaborativeweb}. A classification of the web recommender systems is outlined. Three techniques are mostly used, explicit and implicit profiling and legacy data. The main challenges of the sector are also discussed.

Collaborative filtering is studied in \cite{survey_cft} where the main challenges of the field are also discussed. Furthermore, collaborative filtering techniques are classified to memory-based, model-based and hybrid approaches and the basic techniques of each category are described. The most recent survey oriented to collaborative filtering algorithms is \cite{cacheda}. Various collaborative filtering techniques are described and compared and experiments are performed on MovieLens and Netflix datasets.

A comparison of the performance of the main collaborative filtering algorithms using the MovieLens dataset is given in \cite{Candillier:2007:CSC:1420326.1420378}. The most popular collaborative filtering algorithms are described and their MAE and RMSE is presented, as long as their execution time. This work points out that Bayes models provide an important advantage because of their updatability. Also, applying the K-means clustering algorithm to collaborative filtering gave better results than the usage of other clustering algorithms. Another conclusion of this paper is that item-based recommendations offered better results. 

Since collaborative filtering is one of the most used techniques, recently in \cite{atisha} is realized another survey on this technique. Various collaborative filtering approaches are discussed, mostly emphasizing on how they cope with the most common challenges of collaborative filtering recommendations. This work concludes to the fact that more research is needed to address sparsity issues, for sparsity affects the quality of the recommendations and also because datasets are expected to be even sparser in the future. 

Collaborative filtering techniques are also surveyed in \cite{cfsurvey}. The main concepts of collaborative filtering are presented, providing an overview of the challenges of the field and a brief description of the various methods and the metrics that are used for their evaluation. 
 
A survey that review recent developments in recommender systems and discusses the major challenges is \cite{Lü20121}. The most used algorithms are presented in detail as long as a comparison of their performance measuring MAE and RMSE on the two most preferred datasets, Netflix and MovieLens.

The different algorithms that are used in user-based and item-based techniques are analyzed in \cite{survey5} and the metrics used for evaluation are discussed. A hybrid approach is proposed, which first uses user and item clustering and then the results of both user and item-based algorithms are used to generate the recommendation.

Context-aware technology enhanced recommender systems are discussed in one of the most recent surveys \cite{10.1109/TLT.2012.11}. A classification framework of the context information is introduced, which assigns the contextual information among 8 categories. The existing context-aware recommender systems that are used for technology enhanced learning are analysed concerning the proposed framework. Furthermore, the challenges encountered in the evolution of the field are commented.

Tag-aware recommender systems are surveyed in \cite{Tagsurvey}. Network-based, tensor-based and topic-based models are discussed. The algorithms are evaluated using ranking score, AUC, recall and inter diversity metrics and three different datasets. A comparison is provided of the algorithmic accuracy.

In \cite{Crossdomainsurvey} is proposed a taxonomy for cross-domain recommender systems and a collection of the recent approaches is surveyed.

In \cite{Park:2012:LRC:2181339.2181690} is presented a literature review of the papers concerning recommender systems that have been published in scientific journals during the last decay. The papers are classified according to their publication year, the data mining techniques that they apply and the nature of the recommended items. This work states that the k-nearest neighbour is the most applied data mining technique, especially to collaborative filtering recommender systems.

Recently a study on heterogeneous recommender systems is done in \cite{Bellogín2013142}. The effectiveness of various sources of information, such as tags, social contacts and ratings is investigated, and a variety of content-based, collaborative filtering and social recommender systems is evaluated on Delicious, Last.fm and MovieLens datasets. A comparative evaluation of some social, collaborative filtering and hybrid recommender systems is done in \cite{Bellogin:2013:ECS:2414425.2414439}. Experimental results are analytically presented and discussed in both articles.

The most recent survey on recommender systems is \cite{Bobadilla:2013:RSS:2483330.2483573}. This survey offers an overview of the evolution of the recommender systems' field. Basic terminology as long as an analysis of the most common evaluation metrics are presented. Emphasis is given to the works that exploit social information to provide improved recommendations. This work shows the importance that have the various sources of information to the recommendation process and points out their increasing usage in the future.  
 
A detailed presentation of the field of recommender systems and the most popular techniques that are used, such as collaborative filtering, content-based filtering, data mining and context aware systems are dealed in \cite{RicciHand} and \cite{jannachRS}. Various applications are described and a variety of topics is addressed, such as trust issues and challenges. However, the algorithms' scalability is not covered and no chapter devoted to parallel and distributed applications in the field of recommender systems exist in these books, neither in the above surveys.

\section{Distributed Implementations}
\label{surveydistributed}

In this section distributed implementations of collaborative filtering recommender systems are discussed. The implementations will be classified into the collaborative filtering categories that are analysed in \cite{survey_cft}. The implementations belonging to each category will be discussed according to their chronological appearance. This methodology is followed in order to show how the distributed recommender systems' research evolved through years and  to offer a spherical view of what has been achieved. 

Another factor that will be taken into account is the experiments that have been realized and the metrics that have been preferred for evaluation. Analysing such factors will reveal the most followed methods and will be helpful to the researchers in the future as a reference to conduct experiments that can be easily reproduced and offer results that can be compared to the existing results. Table \ref{table:Distributedcategories} provides a list of all the implementations that are presented in this section.

\begin{table}[htp]
\scriptsize
\begin{center}
\begin {tabular}{|c|c|c|c|}
\hline
\normalsize{Reference} & \normalsize{Year} & \normalsize{Category} & \normalsize{Description}\\
\hline
\cite{Olsson98decentralisedsocial} & 1998 & HYBRID & Content-based, Collaborative and Social Filtering (Item-based) \\
\hline
\cite{Harth} & 2001	& MODEL	& iOwl tool, Association Rules	\\
\hline
\cite{Tveit:2001:PBR:381461.381466}	& 2001	& MEMORY & User-based CF	\\
\hline
\cite{Canny:2002:CFP:829514.830525}	& 2002	& MODEL	& P2P SVD	\\
\hline
\cite{Han2004203}, \cite{Han2004}	& 2004	& MEMORY & User-based CF	\\
\hline
\cite{Ali:2004:TMS:1014052.1014097}	& 2004	& HYBRID & Item-based and Bayesian Content-based Filtering\\
\hline
\cite{MillerKR04} &	2004 & MEMORY	& Item-based\\
\hline
\cite{Berkovsky05collaborativefiltering} & 2005	& MEMORY & Traditional CF User-based	\\
\hline
\cite{Link05distributedrecommender}	& 2005 & HYBRID	& Neighbourhood and Content-based \\
\hline
\cite{AwerbuchPPT05} & 2005 & HYBRID	& Random Product or User Probation	\\
\hline
\cite{Wang:2006:DCF:1141277.1141522} & 2006	& MEMORY & User-item relevance model and Top-N CF (Item-based)\\
\hline
 \cite{Castagnos:2006:CUC:1567016.1567150} 	& 2006	& HYBRID & Hierarchical Clustering and User-based	\\
\hline
\cite{Berkovsky06hierarchicalneighborhood}	& 2006 & MEMORY & Hierarchical formation in the CF algorithm (User-based)	\\
\hline
\cite{Xie20071349} & 2007 & MEMORY & CF with Most Same Opinion and Average Rating Normalization (User-based) \\
\hline
\cite{Berkovsky:2007:EPP:1297231.1297234}	& 2007 & MEMORY	& CF with data obfuscation (User-based)	\\
\hline
\cite{Berkovsky:2007:DCF:1297231.1297238}	& 2007 & MEMORY & CF with domain specialization (Item-based)	\\
\hline
\cite{Castagnos:2007:PCD:1763653.1763695}	& 2007 & MEMORY	& User-based	\\
\hline
 \cite{Ruffo:2009:PRS:1462159.1462163}	& 2009	& MEMORY & Affinity networks User-based	\\
\hline
\cite{Ahn:2010:TFD:1913793.1914138}	& 2010 & MEMORY	& Expert CF	(User-based)\\
\hline
\cite{Isaacman:2011:DRP:2043932.2043948} 	& 2011	& MODEL & Distributed Gradient Descent	\\
\hline
\cite{abs-1109-3318}	& 2011 & MODEL	& User profiling via spectral methods\\
\hline
\cite{kumar} & 2012	& HYBRID 	& Context aware p2p service selection and recovery (CAPSSR)	\\
\hline
\end{tabular}
\caption{List of Distributed Implementations}
\label{table:Distributedcategories}
\end{center}
\end{table}

Recommender systems developed using distributed computing techniques have been initially proposed by \cite{Olsson98decentralisedsocial}, \cite{Harth} and \cite{Tveit:2001:PBR:381461.381466}. In early distributed collaborative filtering recommender system approaches no preference is noticed to any specific algorithmic category.

In \cite{Olsson98decentralisedsocial} a method that combines content-based collaborative filtering and social filtering is proposed. In \cite{Harth} a model-based recommender system named iOwl, that works both as a server and as a client, suggests links to web pages to its users using Association rules. These two approaches propose models that collect data from web sites, thus it will not be available for reuse. As a result, the repetition of any conducted experiments will be hard. A memory-based approach that uses the Pearson correlation coefficient on a peer-to-peer (P2P) architecture similar to Gnutella \cite{Ripeanu02mappingthe} is described in \cite{Tveit:2001:PBR:381461.381466}. The above mentioned approaches emphasize to the description and analysis of the proposed model without conducting any experiments. Therefore, no evaluation is provided. However, those methods consist the opening of the field of distributed recommender systems.

\subsection{Distributed Memory-based Collaborative Filtering}

In this section distributed implementations of memory-based collaborative filtering algorithms are presented. Initially, the traditional user-based and item-based collaborative filtering methods have been chosen for implementation.

In \cite{Han2004203} and \cite{Han2004} the user-based algorithm is implemented on a peer-to-peer architecture through a distributed hash table method. Different parts of the user database are distributed to the peers in such way that all users in the same peer have rated at least one item with the same value. After the similar peers are found, a local training set is constructed and the similar users' vote vectors are retrieved and used to compute the prediction. \cite{MillerKR04} uses five peer-to-peer architectures to examine the item-based algorithm's performance. A model is created for the users while they are online, which is used even if they are offline. In \cite{Berkovsky05collaborativefiltering} the traditional collaborative filtering algorithm is applied over a set of distributed data repositories. Data is distributed both geographically and by topic.

Although in \cite{Han2004203} and \cite{Han2004} different dataset than in \cite{MillerKR04} and \cite{Berkovsky05collaborativefiltering} is used, in all the implementations the MAE metric is used. In \cite{MillerKR04} recall, coverage and memory usage are also measured. It would be interesting to test all the proposed algorithms on the same datasets, in order to compare the prediction accuracy of the different approaches.

Next, more sophisticated ideas that combine the traditional collaborative filtering algorithms with other methods have been developed. In \cite{Wang:2006:DCF:1141277.1141522}, item similarity is calculated by log-based user profiles collected from the Audioscrobbler community \cite{audioscrobbler}. The items are distributed over a peer-to-peer network and the relevance between two items is updated only when an item is downloaded by a peer. The similarities between items are stored locally at item-based tables. Finally, the top-N ranked items are recommended to the user. In \cite{Berkovsky06hierarchicalneighborhood}, a hierarchical neighbourhood is formed, which consists of super-peers and peer-groups. Super-peers are responsible for computations within their peer-group and aggregate their results before notifying them to the active user. In \cite{Xie20071349} is proposed a distributed collaborative filtering algorithm based on the traditional memory-based collaborative filtering. The proposed algorithm locates the similar users using a distributed hash table (DHT) scheme. The number of users that contribute to the recommendation is reduced by using the concept of the Most Same Opinion. Thus, only the ratings of the users with highest consistency with the active user are used. Furthermore, to avoid loosing users who have similar taste but do not rate identically the items, Average Rating Normalization is applied. In \cite{Berkovsky:2007:EPP:1297231.1297234}, a distributed storage of user profiles is combined with data alteration techniques in order to mitigate privacy issues. This approach is focusing on the effect of obfuscating the ratings on the accuracy of the predictions. Domain specialization over the items is developed in \cite{Berkovsky:2007:DCF:1297231.1297238} to confront the data sparsity problem. The ratings matrix is partitioned into smaller matrices that contain ratings given to items that belong to a certain type. In \cite{Wang:2006:DCF:1141277.1141522} is given the coverage and precision of the recommendations. In \cite{Berkovsky06hierarchicalneighborhood}, \cite{Berkovsky:2007:EPP:1297231.1297234},  \cite{Berkovsky:2007:DCF:1297231.1297238} and \cite{Xie20071349} the MAE metric is used and the variety of the datasets used can be seen in table \ref{table:DistrMem}.

A variation of the user-based collaborative filtering algorithm is proposed in \cite{Castagnos:2007:PCD:1763653.1763695}. Each user has his own profile and a single ID. The users can affect the degree of personalization implicitly. The Pearson correlation coefficient is used for the similarity computation and the nearest neighbours of the active user are selected. Four lists of IDs are kept for each user, representing the most similar users, the ones that exceed the minimum correlation threshold, the black-listed users and those that have added the active user to their group profile. Since there is no need to store any neighbours' ratings or similarities, this model has the advantage that it is low memory-consuming. The algorithm is evaluated on the MovieLens dataset, measuring the MAE metric and the computation time.

In \cite{Ruffo:2009:PRS:1462159.1462163} is described a peer-to-peer recommender system that instead of employing users' profiles to produce the recommendations, it uses affinity networks between the users. The affinity networks are generated according to the files that the peers are sharing. In \cite{Ahn:2010:TFD:1913793.1914138} is presented a distributed expert collaborative filtering \cite{Amatriain:2009:WFC:1571941.1572033} recommender system. In expert collaborative filtering the peer user ratings are replaced with ratings provided by domain experts. In this implementation the expert ratings are acquired from \cite{metacritic}. The expert ratings are stored to the server, in a matrix that is used by the clients during the recommendation process. The distributed expert collaborative filtering approach has the advantage that it deals well with privacy issues, since user profiles information is maintained in user's machines.

\begin{table}[htp]
\scriptsize
\begin{center}
\begin {tabular}{|c|c|c|c|c|}
\hline
\normalsize{Ref.} & \normalsize{Algorithm} & \normalsize{Technologies} & \normalsize{Datasets} & \normalsize{Metrics}\\
\hline
\cite{Tveit:2001:PBR:381461.381466} & User-based CF & Java & N/A & N/A  \\
\hline
\cite{Han2004203}  &PipeCF &Distributed &EachMovie & MAE\\ 
 \cite{Han2004} & & Hash Table& & \\ 
\hline
\cite{MillerKR04}  & PocketLens&Chord architecture &MovieLens &Neighborhood similarity \\ 
  & Item-based &for P2P file sharing & &MAE, recall, coverage \\ 
  & & networks& &Memory usage, prediction time \\
\hline
\cite{Berkovsky05collaborativefiltering}  & Traditional CF& Loud Voice Platform& MovieLens&MAE \\ 
\hline
\cite{Wang:2006:DCF:1141277.1141522}  & User-Item & N/A & Audioscrobbler & Coverage\\ 
   & Relevance Model& & & Precision\\ 
\hline
\cite{Berkovsky06hierarchicalneighborhood} &Distributed Hierarchical & Java simulation &MovieLens &MAE \\ 
  &Neighborhood Formation&  &EachMovie & \\ 
  &in the CF algorithm & & Jester & \\ 
\hline
\cite{Xie20071349} & DCFLA & Algorithmic simulation & EachMovie & MAE \\
\hline
 \cite{Berkovsky:2007:EPP:1297231.1297234} &Distributed storage  & Java simulation & MovieLens & MAE\\ 
  &of user profiles& & & \\ 
\hline
 \cite{Berkovsky:2007:DCF:1297231.1297238} & Item Clustering & Java simulation &EachMovie & MAE\\ 
\hline
\cite{Castagnos:2007:PCD:1763653.1763695} &User-based & JXTA&MovieLens &MAE \\
  & AURA&Platform& & Computation time\\  
\hline
 \cite{Ruffo:2009:PRS:1462159.1462163} & Affinity networks&Modification of &self collected &Average \\ 
  & &Phex (Java file sharing ap.) & & accuracy\\ 
\hline
\cite{Ahn:2010:TFD:1913793.1914138} & Expert CF & RIA (Java, & Collected from & N/A \\
 &  & RESTful,XML-RPC)  & metacritic.com, &  \\
 &  &  & rottentomatoes.com &  \\
\hline
\end{tabular}
\caption{Distributed Memory-based Implementations}
\label{table:DistrMem}
\end{center}
\end{table}

\subsection{Distributed Model-based Collaborative Filtering}
\label{surveydistributed1}

In this section the distributed model-based collaborative filtering implementations will be briefly presented. The first distributed recommender system implementation for which an evaluation is provided is \cite{Canny:2002:CFP:829514.830525}, where a peer-to-peer SVD model is proposed. This work is focusing on privacy issues and recommendations are provided from a distributed computation of an aggregate model of users' preferences. Other dimensionality reduction based algorithms that have been implemented in a distributed fashion are briefly described below.
  
Amongst the most popular matrix factorization techniques is the SGD algorithm. A distributed implementation of this algorithm is proposed in \cite{Isaacman:2011:DRP:2043932.2043948}.  In \cite{Isaacman:2011:DRP:2043932.2043948} the information that users provide over items is only available to the users that produced these items. 
 
Another dimensionality reduction algorithm is developed in \cite{abs-1109-3318}. A distributed user profiling algorithm creates a profile vector for each user that represents his taste. Considering a network that is described by an undirected graph, a similarity value is calculated between all the nodes that are connected. The eigenvectors of the adjacency matrix defined from the similarity values are computed in a distributed way and are used to form the recommendations.

The datasets and metrics used in the above implementations can be seen in table \ref{table:DistrModel}.

\begin{table}[htp]
\scriptsize
\begin{center}
\begin {tabular}{|c|c|c|c|c|}
\hline
\normalsize{Ref.} & \normalsize{Algorithm} & \normalsize{Technologies} & \normalsize{Datasets} & \normalsize{Metrics}\\
\hline
\cite{Harth} & Association Rules & Python, iOwl & N/A & N/A \\
\hline
\cite{Canny:2002:CFP:829514.830525}   & P2P SVD&Matlab &EachMovie & MAE\\ 
  & & & & Average recommendation\\ 
 & & & & time \\
\hline
 \cite{Isaacman:2011:DRP:2043932.2043948}  & Distributed Gradient & Facebook ap.& Netflix & RMSE \\ 
   & Descent&WebDose & & Probability distribution\\ 
  & & & &Estimation of rating \\ 
\hline
\cite{abs-1109-3318}  &Similarity-based & Mathematical simulation & Netflix (synthetic) &Convergence of the \\ 
  &profiling & & &asynchronous distributed \\ 
  & & & & algorithm \\ 
\hline
\end{tabular}
\caption{Distributed Model-based Implementations}
\label{table:DistrModel}
\end{center}
\end{table}

\subsection{Hybrid Distributed Collaborative Filtering Methods}

Except from \cite{Olsson98decentralisedsocial}, a few more hybrid distributed methods have been developed. These implementations can be seen in table \ref{table:DistrHybrid}.

In \cite{Ali:2004:TMS:1014052.1014097} a client-server architecture is followed, where item correlations are computed at the server side and are used by the client side to make the predictions. No evaluation of the model is provided.

In \cite{Link05distributedrecommender} is combined memory-based collaborative filtering using neighbours and content-based collaborative filtering. The 'mailing list' model and the 'word-of-mouth' model are described. Users share information with their neighbours according to one of the two models. The intention of the distributed recommender systems that are described in this paper is to notify item information to as many users as possible, that are expected to have an interest in the items. Unfortunately, no details are given on the implementation and its performance needs to be evaluated.

In \cite{AwerbuchPPT05} is described a peer to peer distributed algorithm that focuses on the minimization of the recommendation complexity by avoiding the evaluations provided by the untrusted users. However, the algorithm is only described theoretically and is not implemented.

User-based collaborative filtering employing the Pearson correlation coefficient is combined with a hierarchical clustering algorithm in \cite{Castagnos:2006:CUC:1567016.1567150}. The users' profiles are sent to the server and the system creates virtual communities using the hierarchical clustering algorithm. On the client side takes place the classification of the active user to a group. The predictions are made according to the distances between the active user and the closest group's users. 

In \cite{kumar} is proposed an algorithm for context aware P2P service selection (CAPSSR). Users can access various services available on internet. After using one service, its rating is increased or decreased depending on whether the use of the service was successful or not. For the evaluation of the algorithm the MovieLens and the Jester datasets are used. Scalability, accuracy, efficiency and mean waiting time are evaluated.

\begin{table}[htp]
\scriptsize
\begin{center}
\begin {tabular}{|c|c|c|c|c|}
\hline
\normalsize{Ref.} & \normalsize{Algorithm} & \normalsize{Technologies} & \normalsize{Datasets} & \normalsize{Metrics}\\
\hline
 \cite{Olsson98decentralisedsocial} & Content-based filtering  & Agent-based & N/A &N/A \\
  &   CF and Social filtering &  &  & \\
 \hline
 \cite{Ali:2004:TMS:1014052.1014097} & Item-based & Proprietary & Tivo data & N/A \\
 & Bayesian content-based  &  &  &  \\
  &  filtering &  &  &  \\
\hline
\cite{Link05distributedrecommender} & User Neighbourhood and & Mathematical  & N/A &N/A \\
 & Content-based Filtering &simulation & & \\
\hline 
 \cite{Castagnos:2006:CUC:1567016.1567150} &User-based &Java & MovieLens& MAE\\ 
  & Hierarchical clustering& & &Computation time \\ 
\hline
\cite{AwerbuchPPT05} & Random product or  & Mathematical  & N/A & N/A\\
 &  user probation &  simulation &  & \\
\hline
 \cite{kumar}   &Context Aware & N/A & MovieLens& Scalability \\ 
  & P2P Service& & Jester& Accuracy \\
  & CAPSSR& & &DFM \\
& & & & Mean waiting time\\
& & & & Precision\\
\hline
\end{tabular}
\caption{Distributed Hybrid Implementations}
\label{table:DistrHybrid}
\end{center}
\end{table}

\section{Parallel Implementations}
\label{parallel}

\subsection{Distributed Memory Implementations}
\label{cluster}

This section presents the parallel implementations that are built on distributed memory systems. A list of these approaches is provided in table \ref{table:Parallelcategories} and more information can be seen in table \ref{table:ParallelModelCluster}. As can be seen in these tables, no memory-based algorithms are implemented on distributed memory systems and a clear preference is noticed to the model-based algorithms. In this section the implementations are presented according to the implemented algorithm.

\begin{table}[htp]
\scriptsize
\begin{center}
\begin {tabular}{|c|c|c|c|}
\hline
\normalsize{Reference} & \normalsize{Year} & \normalsize{Category} & \normalsize{Description}\\
\hline
\cite{co_cluster_2005}	& 2005 & MODEL	& Bregman Co-clustering\\
\hline
\cite{Par_cf_netflix} 	& 2008	& MODEL & ALS-WR	\\
\hline
\cite{Chen:2008:CCF:1401890.1401909}	& 2008	& HYBRID	& Combinational CF	\\
\hline
\cite{kwon} 	&2010	& MODEL & Bregman Co-clustering	\\
\hline
\cite{Liu:2011:PPL:1961189.1961198} & 2011 & MODEL & PLDA+ \\
\hline
\cite{mfrm} & 2012 & MODEL& Coordinate Descent CCD++ \\
\hline
\cite{dmcomp} & 2012 & MODEL & DALS, ASGD, DSGD++ \\
\hline
\cite{6413871} & 2012 & MODEL & Co-clustering \\
\hline
\end{tabular}
\caption{List of Implementations on Distributed-memory Systems}
\label{table:Parallelcategories}
\end{center}
\end{table}

Clustering is a very often used model-based collaborative filtering method. In \cite{co_cluster_2005} and \cite{kwon} the Bregman co-clustering algorithm \cite{Banerjee_Dhillon_Ghosh_Merugu_Modha_2004} is parallelized. In \cite{co_cluster_2005} user and item neighborhoods are simultaneously created by dividing among the processors submatrices of the rows and colums of the ratings matrix. A comparison of the proposed algorithm with SVD \cite{Sarwar_Karypis_Konstan_Riedl_2000}, NNMF \cite{Hofmann_2004} and classic correlation-based filtering \cite{Resnick_Iacovou_Suchak_Bergstrom_Riedl_1994} is provided. In \cite{kwon} the row and column cluster assignments are performed in parallel, by also dividing the rows and columns among processors. In both implementations MPI is used.

Another co-clustering based collaborative filtering algorithm is proposed and examined in \cite{6413871}. The algorithm's performance is compared to the authors' previous work \cite{IBM_India_11}. The initial ratings matrix is partitioned according to a certain number of rows and columns, and to each partition is applied the algorithm described in \cite{IBM_India_11}. The row and column clusters formed in each partition are merged with the neighbouring partition. This procedure is followed to various levels of row and column clusters until the whole matrix is obtained as a single partition. Then the flat parallel co-clustering runs once more. This hierarchical co-clustering algorithm aims in achieving a reduced communication and computation cost. The performance of the proposed algorithm is examined through the Netflix and Yahoo KDD Cup datasets. The experiments are conducted on the Blue gene/P architecture and RMSE is the accuracy  metric used. Detailed scalability analysis is also provided. 

A distributed LDA algorithm is described in \cite{Liu:2011:PPL:1961189.1961198} and is implemented using MPI. This implementation improves the scalability of the author's previous effort \cite{Wang:2009:PPL:1574036.1574062} and reduces the communication time by applying methods such as data placement, pipeline processing, word bundling and priority-based scheduling.

In \cite{Par_cf_netflix} the Alternating Least Squares with Weighted λ Regularization algorithm (ALS-WR) is implemented using parallel Matlab. The updates of U and M matrices are parallelized and the rows and columns of the ratings matrix are distributed over the cores.
 
The ALS and SGD algorithms that are used for matrix factorization are parallelized in \cite{dmcomp}. The parallel ALS (PALS), parallel SGD (PSGD), distributed ALS (DALS), asynchronous SGD (ASGD) and DSGD-MR along with its extension DSGD++, are described, implemented and compared. All the above algorithms are implemented in C++ and for communication over the nodes of the distributed algorithms MPICH2 is used. The Netflix dataset and the dataset of Track 1 of the KDD Cup 2011 contest are used. The time an iteration needs to be completed, the number of iterations required to converge and the total time to converge of the algorithms are compared.

In \cite{mfrm} a coordinate descent algorithm is proposed , CCD++ that approximates the ratings matrix by $WH^{T}$, updating one variable at a time, while maintaining the other variables fixed. The algorithm is parallelized on a MPI cluster. Each machine updates different subvectors of the row vectors of $W$ and $H$ and broadcasts the results. The CCD++, ALS and SGD algorithms are parallelized and compared. The training time and the speedup are measured. MovieLens 10M, Netflix and Yahoo! Music datasets are used for the experiments.

In \cite{Chen:2008:CCF:1401890.1401909} a collaborative filtering method for community recommendation for social networking sites is proposed. Parallel Gibbs sampling and parallel Expectation Maximization algorithm are combined. Experiments are performed on the Orkut dataset, measuring the implementation's speedup. Furthermore, an analysis of the  computation and communication time is provided. However, no information is given on the technologies used to achieve the algorithm's parallelization.

\begin{table}[htp]
\scriptsize
\begin{center}
\begin {tabular}{|c|c|c|c|c|}
\hline
\normalsize{Ref.} & \normalsize{Algorithm} & \normalsize{Technologies} & \normalsize{Datasets} & \normalsize{Metrics}\\
\hline
\cite{co_cluster_2005} &Parallel  &C++, MPI & MovieLens& MAE, Average prediction time \\ 
  &Co-clustering &LAPACK Library &Bookcrossing &Training time \\ 
  &Bregman & & & Comparison to SVD,NNMF \\ 
  & & & &and classic correlation-based filtering \\ 
\hline
\cite{Par_cf_netflix}   &ALS-WR &Parallel Matlab, & Netflix & RMSE \\ 
  & &MPI &  &  \\ 
\hline
\cite{Chen:2008:CCF:1401890.1401909}&Combinational & MPI & Orkut& Speedup, Computation/\\ 
  & CF (CCF)& & (synthetic) & communication time \\
  & & & & analysis\\
\hline
\cite{kwon}  &Bregman &MPI &Netflix &Speedup \\
  &Co-clustering  & & &Time per iteration \\  
\hline
\cite{Liu:2011:PPL:1961189.1961198} & PLDA+ & MPI & NIPS,  &Speedup,\\ 
& & & Wiki 20T,  & Communication time,\\
& & & Wiki 200T & Sampling time\\
\hline
\cite{mfrm}  &Coordinate & C++ and MPI&MovieLens & Speedup,\\ 
  &Descent CCD++ &  &Netflix & Training time\\
  & & &Yahoo! music & \\
\hline
\cite{dmcomp} & DALS, ASGD & C++ & Netflix &Time per iteration, \\
  & DSGD++ &MPICH2 &KDD Cup 2011 &Number of iterations, \\
  & & &(Track 1) &Total time to converge \\
\hline

 \cite{6413871} & Co-clustering& MPI&Netflix &RMSE \\ 
 & & & Yahoo KDD Cup& Speedup\\
\hline
\end{tabular}
\caption{Parallel Implementations on Distributed Memory Environments}
\label{table:ParallelModelCluster}
\end{center}
\end{table}

\subsection{Shared Memory Implementations}
\label{multithreaded}

Recommendation algorithms that have been implemented on shared memory architectures will be discussed in the present section. A list of these implementations is given in table \ref{table:Multithreadedcategories}.

In \cite{IBM_India_10} is presented a parallel model-based collaborative filtering algorithm based on the Concept Decomposition technique for matrix approximation. This technique performs clustering with the k-Means algorithm and afterwards solve a least-squares problem. The proposed algorithm consists of four multithreaded stages, concluding to the prediction phase. Posix Threads are used to implement the proposed method, which is evaluated on the Netflix dataset. Training and prediction time are measured as long as the RMSE metric. A detailed scalability analysis is also presented

Parallel Gradient Descent in a shared memory environment is approached in \cite{LOU1}. In this approach, if the parameter $\theta$ is already processed, the other processors skip the update and the processor with the most queued updates is the next processor that gains access to update $\theta$. This method is aiming to reduce the idle time of the processors.

In \cite{recht} an incremental SGD is implemented on multicore processors. One core is assigned for the ordering and partitioning of the data into chunks. Non-overlapping chunks are grouped into rounds and each round's chunks are accessed by a different process.

In \cite{RechtRWN11} SGD is implemented without locking the access to shared memory. Memory overwrites are not avoided, but they are very rare because of data sparseness. Therefore, they don't cause errors to the computations.

In \cite{karydi_margaritis_1} is described a multithreaded application of the memory-based Slope One algorithm, implemented with the OpenMP Library. Each thread assumes the computations on a different part of the ratings matrix. The MovieLens dataset is used for the performance and scalability evaluation and the metrics used for the evaluation can be seen in table \ref{table:multithreaded}.

The CCD++ algorithm \cite{mfrm} described in section \ref{surveydistributed1} is also parallelized on a multi-core system using the OpenMP library. Each core updates different subvectors of the row vectors of $W$ and $H$. Parallel implementations of the CCD++, ALS and SGD algorithms are compared by means of the running time against RMSE and speedup. The datasets used for the experiments can be seen in table \ref{table:multithreaded}.

A new parallel matrix factorization approach based on SGD is analysed in \cite{Zhuang:2013:FPS:2507157.2507164}. The FPSGD method is designed for shared memory systems and embodies two techniques. Lock-free scheduling to avoid data imbalance and partial random method to address memory discontinuity. A comparison among other parallel SGD methods (\cite{Gemulla}, \cite{RechtRWN11} and \cite{mfrm}) is provided and after applying optimizations such as cache-miss reduction and load balancing, FPSGD is proved more efficient. Information is given on the algorithm's run time and RMSE is used to evaluate the implementation. The MovieLens, Netflix and Yahoo!Music datasets are used for the experiments.

\begin{table}[htp]
\scriptsize
\begin{center}
\begin {tabular}{|c|c|c|c|}
\hline
\normalsize{Reference} & \normalsize{Year} & \normalsize{Category} & \normalsize{Description}\\
\hline
\cite{IBM_India_10} & 2010 & MODEL & Concept Decomposition\\
\hline
\cite{LOU1} & 2010 & MODEL & Asynchronous Gradient Descent \\
\hline
\cite{recht} & 2011	& MODEL & SGD	\\
\hline
 \cite{RechtRWN11}	& 2011	&	MODEL & SGD	\\
\hline
\cite{karydi_margaritis_1} & 2012	& MEMORY	& Slope One\\
\hline
\cite{mfrm} & 2012	& MODEL	&  Coordinate Descent CCD++ \\
\hline
\cite{Zhuang:2013:FPS:2507157.2507164}	& 2013	& MODEL & FPSGD	\\
\hline
\end{tabular}
\caption{List of Implementations on Shared-memory Systems}
\label{table:Multithreadedcategories}
\end{center}
\end{table}

\begin{table}[htp]
\scriptsize
\begin{center}
\begin {tabular}{|c|c|c|c|c|}
\hline
Ref. & Algorithm & Technologies & Datasets & Metrics\\
\hline
 \cite{IBM_India_10} &Concept  & Posix Threads& Netflix & RMSE, Scalability \\ 
  &Decomposition & & &Prediction/training time \\ 
\hline
\cite{LOU1} & Asynchronous Gradient Descent &  N/A & Netflix &  Speedup, Parallel Efficiency \\
 &  & & & RMSE, Wall clock time \\
\hline
\cite{recht}  &Parallel SGD & N/A &MovieLens &Total CPU time \\
  &JELLYFISH & & Netflix& RMSE\\
\hline
 \cite{RechtRWN11}&Multicore SGD & C++&Reuters RCV1 &Speedup \\ 
  &HogWild!& &Netflix & \\ 
  & & &KDD Cup 2011 (Task 2) & \\ 
  & & &Jumbo (synthetic) & \\ 
  & & & Abdomen& \\ 
\hline
 \cite{karydi_margaritis_1} &Slope One & OpenMP &MovieLens &Scalability, Speedup \\ 
  & & & & Total elapsed time \\ 
  & & & &Prediction per second \\ 
  & & & &Prediction  time per rating \\ 
\hline
\cite{mfrm}  &Coordinate & C++ and OpenMP&MovieLens &Running time vs RMSE, \\ 
  &Descent CCD++ &  &Netflix &Speedup,\\
  & & &Yahoo!Music & \\
\hline
\cite{Zhuang:2013:FPS:2507157.2507164}  &FPSGD & C++ &MovieLens &Total time \\
  & & SSE Instructions & Netflix & RMSE\\
    & & & Yahoo!Music & \\
\hline
\end{tabular}
\caption{Imlementations on Shared-memory Systems}
\label{table:multithreaded}
\end{center}
\end{table}

\subsection{GPU-based Implementations}
\label{surveygpu}

Recently general purpose computations on GPU devices have emerged as an attractive solution for parallel computing. The performance of implementations belonging to various areas of computer science has been significantly increased when GPUs are used. This section presents implementations of collaborative filtering algorithms that have been parallelized on GPU devices. First the memory-based implementations will be described according to their chronological appearance and afterwards the model-based approaches will be discussed according to the algorithm they implement. In table \ref{table:GPUcategories} can be seen a list of all the implementations on GPU that will be discussed above.

\begin{table}[htp]
\scriptsize
\begin{center}
\begin {tabular}{|c|c|c|c|}
\hline
\normalsize{Reference} & \normalsize{Year} & \normalsize{Category} & \normalsize{Description}\\
\hline
\cite{Bondhugula} & 2006 & MODEL & SVD \\
\hline
\cite{5161058} & 2009 & MODEL & SVD \\
\hline
\cite{katosvdgpu} & 2010 & MODEL & SVD \\
\hline
\cite{Kato:2010:SKN:1844765.1845125} &  2010 & MEMORY & K-nearest neighbor \\
\hline
\cite{6064611} & 2011 & MODEL & Co-clustering\\
\hline
 \cite{Li:2011:SNT:1998076.1998131} & 2011 & MEMORY & Top-N user-based random walk\\
\hline
\cite{6337114} & 2012 & MEMORY & Item-based CF user-based CF\\
\hline
\cite{ChuaGpuR} & 2012 & MODEL & Approximate SVD\\
\hline
\cite{Cai:2012:GRB:2403514.2403537} & 2012 & MODEL & RBM-CF \\
\hline
\cite{Zastrau:2012:SGD:2406479.2406497} & 2012 & MODEL &  SGD\\
\hline
\cite{6384989} & 2012 & MEMORY  & User-based CF\\
\hline
\cite{Foster:2011:GAS:2351958.2352024} & 2012 & MODEL & Aproximate SVD \\
\hline
\cite{GPU14} & 2013 & MODEL & RBM-CF \\
\hline
\end{tabular}
\caption{List of Implementations on GPUs}
\label{table:GPUcategories}
\end{center}
\end{table}

\subsubsection{Memory-based Implementations on GPU.}

The k-nearest neighbour problem is confronted in \cite{Kato:2010:SKN:1844765.1845125}, where an algorithm is introduced that finds the $k$ most similar users using GPUs. The Hellinger distance is employed and the algorithm is implemented in CUDA. The problem of computing the distances is divided into blocks which are called grids. Each GPU processes a grid. Each grid is divided row-wisely in blocks, which are assigned to thread blocks. Each thread assumes a row of the block. For the selection of the nearest neighbours, the threads in a block simultaneously process their corresponding part of data and realize the necessary computations. 

In \cite{Li:2011:SNT:1998076.1998131} is described a hybrid parallel top-N recommendation algorithm that aims to face the cold-start user problem and the scalability problem. The proposed algorithm combines user-based collaborative filtering with random walk on trust network and merges the results to provide the top-N recommended items. First runs the user-based algorithm where the similarities between users are computed by Pearson correlation. A heap structure is used to help selecting a subset of similar users. Finally, random walks are used to define a subset of trusted users. The results brought by the two algorithms are merged to provide the final top-N recommendations. All three parts of the algorithm are implemented in CUDA. 

The traditional item and user-based collaborative filtering algorithms are parallelized in \cite{6337114}. The performance of the proposed algorithms is examined under Intel's Single Chip Cloud Computer (SCC) and under NVIDIA's Cuda-enabled GPGPU co-processor. The similarity measure used is the Pearson correlation coefficient. The identification of common items is usually achieved by means of brute force methods. This approach avoids such methods by using an intermediate matrix. The number of co-rated items is calculated and subsequently the intermediate matrix is used to calculate the correlation coefficient.


Another implementation of the user-based Collaborative Filtering algorithm on GPU is approached in \cite{6384989}. Three different approaches are investigated. First the Pearson correlation coefficient is used. Afterwards, implied similarities are calculated. Implied similarity is based on the common neighbours among users. Finally, the empty cells of the ratings matrix are filled with the value of the average rating for each user. The accuracy of the three approaches, as long as the total execution time on both CPU and GPU are examined using a part of a dataset provided by GroupLens.

Table \ref{table:MemoryGPU} shows the datasets on which the above implementations conduct experiments and the metrics used for evaluation.

\begin{table}[htp]
\scriptsize
\begin{center}
\begin {tabular}{|c|c|c|c|c|}
\hline
Ref. & Algorithm & Technologies & Datasets & Metrics\\
\hline
\cite{Kato:2010:SKN:1844765.1845125}  &K-nearest &CUDA & N/A & Total elapsed time \\ 
 & Neighbor& & &  \\ 
\hline
\cite{Li:2011:SNT:1998076.1998131} &Top-N &C++, &Flixster & Recall \\ 
& User-based CF& CUDA& &Speedup  \\ 
& Random walk& & &  \\ 
\hline
\cite{6337114} &User-based &CUDA & Flixter (synthetic)&   Execution time \\ 
&Item-based & &Bookcrossing (Subset) &  Power/energy consumption \\ 
& & & MovieLens (Subset)& Speedup\\
\hline
\cite{6384989}&User-based &CUDA &GroupLens(Subset) &RMSE, Execution time  \\ 
& & & & CPU/GPU time usage \\ 
\hline
\end{tabular}
\caption{Memory-based Implementations on GPU}
\label{table:MemoryGPU}
\end{center}
\end{table}

\subsubsection{Model-based Implementations on GPU.}

Model-based Collaborative Filtering implementations on GPU commenced with an approach to the SVD algorithm \cite{Bondhugula}. First a bidiagonalization of the ratings matrix takes place and then the bidiagonal matrix is diagonalized by implicit-shifted QR algorithm. The diagonalization is performed on CPU. The time needed for the bidiagonalization according to the size of the matrix is measured. Information on how the parallelization on the GPU is achieved is not specified neither is given any information on the used dataset.

Among the first implementations of SVD on GPU is that described in \cite{5161058}. The CUDA architecture and CUBLAS library are used. All the necessary data to perform the bidiagonalization are stored in the GPU memory in order to avoid data transfer between CPU and GPU. The diagonalization of the bidiagonal matrix is also performed on the GPU. The rows of the matrix are divided into blocks and each element of the block is processed by a different thread. The performance is compared to that of an optimized CPU implementation on Matlab and to Intel MKL. Random dense matrices are used for the experiments and the average execution time and speedup are examined. 

In \cite{katosvdgpu} is proposed another parallel version of the SVD on GPU implemented in CUDA. The order of the computations of the U and V matrices is altered. Instead of examining all of the input data step by step, when the element $a_{ij}$ of the sparse matrix A that contains the ratings is processed, the i-th row of U and the j-th row of V are updated. This means that all the rows of U can be updated in parallel. First U is updated for each $a_{ij}\neq 0$ and then V. The results are compared to those of a single threaded implementation on a recent CPU. The time needed for one step of the iteration of convergence is measured.

Approximate Singular Value Decomposition is parallelized in \cite{ChuaGpuR} using R and C languages and CUDA architecture. A single node GPU kernel and a distributed GPU kernel over 6 nodes are used to approximate the matrix A, which contains the ratings. The algorithm is parallelized following the description of \cite{katosvdgpu}. The total execution time and computation versus communication time are given. However, the author reports that the implemented algorithm’s performance is very sensitive to changes in the learning parameters, and only works for square matrices of sizes up to 1024.

Approximate SVD using CUDA is also addressed in \cite{Foster:2011:GAS:2351958.2352024}. The proposed method is based on a SVD method, called QUIC-SVD \cite{Holmes} that is an approximate SVD algorithm that utilizes a tree-based structure. The algorithm is implemented on CUDA architecture with CULA library for linear algebra. Measures have been taken in order to be able to process matrices of larger size than that of the GPU or main memory. The ratings matrix is divided in submatrices and QUIC-SVD runs on every submatrix. Blocks of the ratings matrix are loaded into memory and are sequentially processed. A cosine tree is created for each submatrix and a common basis is shared among the trees. The algorithm's results are compared to those of a multithreaded CPU version and other two implementations of SVD. Random matrices of various sizes are used for the experiments and running time and speedup is provided.

In \cite{6064611} is described the parallelization on the GPU of the non-parametric co-clustering model. In this implementation, computations are made on both CPU and GPU. The speedup of the GPU computations over the CPU computations is measured. Two datasets are used for the collaborative filtering domain, the Netflix dataset and a Facebook dataset of user application consumption.

The Stochastic Gradient Descent algorithm is parallelized on GPU in \cite{Zastrau:2012:SGD:2406479.2406497}. A hush function is created to help in executing threads in parallel. The implementation of the SGD algorithm is compared to an implementation of ALS on GPU and to a batch gradient descent. The Netflix dataset is used and the RMSE is measured as long as execution time and scalability.

One of the main reasons that Restricted Boltzmann Machines are often used to collaborative filtering is their property to easily handle large datasets \cite{Salakhutdinov:2007:RBM:1273496.1273596}. A preference is recently shown to the usage of Restricted Boltzmann Machines for collaborative filtering algorithms on GPUs. A Restricted Boltzmann Machine is applied to collaborative filtering in \cite{Cai:2012:GRB:2403514.2403537} and a parallel implementation on GPU using CUDA is discussed. The computations of the collaborative filtering RBM are remodeled to matrix operations in order to  be implemented in CUDA. The Java programming language and the JCUDA library are used. Experiments run on the Netflix dataset and the implementation’s speedup is examined.

The same authors also applied Restricted Boltzmann Machine on GPU in \cite{GPU14}. The matrix multiplications on GPU described in their previous work \cite{Cai:2012:GRB:2403514.2403537} are adjusted to a hybrid framework that schedules the use of CPU and GPU. A CPU thread controls the scheduler and another thread activates the CUDA kernels. The rest of the CPU cores undertake the multi-processor kernels. The framework is implemented in JAVA and the JCUDA library is used for the CUDA kernels. The speedup of the hybrid implementation is compared to that of a CUDA implementation and to that of a multithreaded implementation. The run time of the hybrid kernel is given and the proportion of the CPU computation and hybrid kernel's run time is discussed. Information about the dataset used is vague.

The technologies and the datasets used by each model-based algorithm implementation on GPUs can be seen in table \ref{table:MOdelGPU}.

\begin{table}[htp]
\scriptsize
\begin{center}
\begin {tabular}{|c|c|c|c|c|}
\hline
Ref. & Algorithm & Technologies & Datasets & Metrics\\
\hline
\cite{Bondhugula}& SVD& CUDA& N/A & Time for bidiagonalization \\ 
& & Intel MKL & & \\ 
\hline
\cite{5161058}&SVD & CUDA& Random dense& Average execution time \\ 
& &CUBLAS Library  &matrices & Speedup \\ 
& &Matlab & & \\
\hline
\cite{katosvdgpu}  & SVD&CUDA & Random data &Time for one step of the  \\ 
& & & &iteration of convergence   \\ 
\hline
\cite{6064611} &Non-parametric & CUDA&Netflix &Speedup  \\ 
 & Co-clustering& & Facebook&AUC  \\ 
\hline
\cite{ChuaGpuR} &Approximate  &R,C& N/A &Total execution time  \\ 
 &SVD&CUDA & &Computation/communication time  \\ 
\hline
\cite{Cai:2012:GRB:2403514.2403537} &RBM for CF &CUDA, Java & Netflix&Speedup  \\ 
& & JCUDA Library& &  \\ 
\hline
\cite{Zastrau:2012:SGD:2406479.2406497} & SGD&CUDA &Netflix & RMSE, Execution time \\ 
 & & & &Speedup  \\ 
\hline
\cite{Foster:2011:GAS:2351958.2352024}& Approximate SVD&CUDA & Random & Running time \\ 
&QUIC-SVD& CULA Library &matrices &  Speedup\\ 
\hline
\cite{GPU14} &RBM for CF &CUDA, Java & Self-generated&Speedup   \\ 
& & JCUDA Library& &Runtime  \\ 
\hline
\end{tabular}
\caption{Model-based Implementations on GPU}
\label{table:MOdelGPU}
\end{center}
\end{table}

\section{Platform-based Recommendations}
\label{surveyframeworks}

Since the available amount of data is continuously increasing, it is inevitable not to consider new methods to facilitate and expedite its elaboration. To this effort, the usage of Big-data frameworks has a significant contribution. This section is devoted to the implementations of collaborative filtering recommendation algorithms realized with the aid of frameworks that are adequate for parallel processing and for handling of large datasets. The implementations will be classified to memory and model-based and they will be discussed according to their publication year, commencing with the oldest one. Table \ref{table:Frameworkcategories} lists the implementations that are based on frameworks.

The field opens with a hybrid approach that provides recommendations to the Google News users \cite{Das:2007:GNP:1242572.1242610}. The model-based PLSI and MinHash clustering algorithms are combined with the item co-visitation counts. The MapReduce framework is used to parallelize the MinHash clustering method and the EM (Expectation Maximization) algorithm. The users' click history constitutes the input of the algorithm's Map phase, which is conducted over various machines. The algorithm outputs key-value pairs that correspond to the clusters that each user belongs to. A comparison of the MinHash and PLSI algorithms proves that their combination performs worst than the original algorithms. Information on the used datasets and the metrics selected for evaluation is provided at table \ref{table:HybFrameworks}. In table \ref{table:MemFrameworks} can be seen the datasets and metrics that are used to each memory-based implementation, and in table \ref{table:ModFrameworks} is given information for the model-based implementations.

\begin{table}[htp]
\scriptsize
\begin{center}
\begin {tabular}{|c|c|c|c|}
\hline
\normalsize{Reference} & \normalsize{Year} & \normalsize{Category} & \normalsize{Description}\\
\hline
\cite{Das:2007:GNP:1242572.1242610} & 2007 & HYBRID & MinHash and PLSI clustering \\
 &  &  & Covisitation counts \\
\hline
\cite{Chen:2009:CFO:1526709.1526801} & 2009 & MODEL & LDA\\
\hline
\cite{Wang:2009:PPL:1574036.1574062} & 2009 & MODEL & PLDA\\
\hline
\cite{co_cluster_netflix} & 2009 & MODEL & Bregman Co-clustering \\
\hline
\cite{Ub_Hadoop} & 2010 & MEMORY & User-based \\
\hline
\cite{5693295} & 2010 & MEMORY & User profiling \\
\hline
\cite{Davidson:2010:YVR:1864708.1864770} & 2010 & MEMORY & Distributed item-based \\
\hline
\cite{NIPS2010_1162} & 2010 & MODEL & SGD\\
\hline
\cite{CF_Hadoop} & 2011 & MEMORY & Item-based \\
\hline
\cite{Gemulla} & 2011 & MODEL & DSGD \\
\hline
\cite{pcf_streaming_data} & 2011 & MODEL & Distributed SGD \\
\hline
\cite{multicore_cf} & 2011 & MEMORY AND MODEL & CF Library: \\
& & & ALS, Wals, BPTF, SGD, \\
& & & SVD++, Item-kNN,\\
& & & Time-kNN, Time-SGD,\\
& & & Time-SVD++, MFITR\\
\hline
\cite{chen} & 2011 & MEMORY & User-based Clustering \\
& & & Slope One  (CWSO) \\
\hline
\cite{Schelter:2012:SSN:2365952.2365984} & 2012 & MEMORY & Pairwise Item Comparison\\ 
& &  & Top-N Recommendation\\
\hline
\cite{Kanagal:2012:SRS:2336664.2336669} & 2012 & MODEL & Taxonomy-aware Latent Factor\\
\hline
\cite{Schelter:2013:DMF:2507157.2507195} & 2013 & MODEL & ALS \\
\hline
\cite{Tang:2013:SMF:2487788.2487801} &  2013 & MODEL & Truncated SVD and ALS \\
\hline
\end{tabular}
\caption{List of Implementations on Frameworks}
\label{table:Frameworkcategories}
\end{center}
\end{table}

\begin{table}[htp]
\scriptsize
\begin{center}
\begin {tabular}{|c|c|c|c|c|}
\hline
Ref. & Algorithm & Technologies & Datasets & Metrics\\
\hline
  \cite{Das:2007:GNP:1242572.1242610}  &MinHash clustering &MapReduce &MovieLens, & Precision, Recall, \\ 
  &EM, PLSI & & GoogleNews& Live traffic ratios \\ 
\hline
\end{tabular}
\caption{Hybrid Implementations on Frameworks}
\label{table:HybFrameworks}
\end{center}
\end{table}

\subsection{Memory-based Implementations}

In \cite{Ub_Hadoop} is implemented a user-based collaborative filtering algorithm following the MapReduce model on the Hadoop platform. The algorithm is divided into three phases. The data partitioning phase, where the user ID's are separated into different files and are used as input during the map phase. The map phase, where the recommendation list for each user is calculated, and the reduce phase, where all information calculated is collected and output is generated. The algorithm's speedup is considered on the Netflix dataset.

A parallel user profiling approach is proposed in \cite{5693295}. The suggested implementation is developed on the Hadoop Map-Reduce framework and Cascading \cite{Cascading}, using the Del.icio.us dataset \cite{delicious} on the Amazon EC2 EMR clouds. In order to create the user profiles a tag vector is formed for each user. The recommendation is obtained by the user-based algorithm, using cosine similarity to select the K-nearest neighbours. The top-N items are recommended according to the prediction value. Three Cascading flows implement the user profiling phase, the formation of the neighbourhood and the recommendation phase. A comparison is given, of the three jobs' running time on the cloud and on a local desktop machine.

Personalized video recommendations are made through YouTube's distributed item-based recommendation system \cite{Davidson:2010:YVR:1864708.1864770}. Item similarity is calculated considering the user's co-visitation counts. In order to process large amounts of data, recommendations are calculated following a batch-oriented pre-computation approach of MapReduce computations. The recommendation quality is evaluated through the following metrics: click through rate (CTR), long CTR, session length, time until first long watch, and recommendation coverage. Unfortunately no other implementation assumes these metrics.

The item-based collaborative filtering algorithm is implemented on Hadoop in \cite{CF_Hadoop}. This approach separates the three most excessi 10.0.4 LAPACKve computations into four Map-Reduce phases, which are executed in parallel on a three node Hadoop cluster. In the first Map-Reduce phase the average rating for each item is computed, in the second  Map-Reduce phase the similarity between item pairs is computed, in the third  Map-Reduce phase the similarity matrix is recorded, and finally the computations for the items prediction take place in the fourth Map-Reduce phase.  The MovieLens dataset is used, and isoefficiency and speedup scalability metrics are used to measure the implementation's performance.

In \cite{chen} is implemented a user-based clustering weighted Slope One (CWSO) algorithm using Hadoop on a 5 machines cluster. This approach clusters users and assigns weights to each cluster. Then, the ratings are predicted using weighted Slope One. The prediction is accomplished with two Map-Reduce phases. To the first phase a list of the items that are rated and belong to the same cluster with the active user's clusters is constructed. To the second phase the average deviation between two items is calculated and used for the prediction. Users are clustered with the K-Means algorithm on WEKA \cite{weka}.  The MovieLens dataset is used, and MAE and accuracy are measured.

A neighbourhood-based algorithm for batch recommendation is implemented on Hadoop MapReduce framework in \cite{Schelter:2012:SSN:2365952.2365984}. One MapReduce phase counts the item coocurencies without taking into account the rating values that have been given to the items. The item vectors are preprocessed in order to compute their norm and their dot products and finally proceed to the similarity computation. Another MapReduce phase applies a threshold to sparsify the similarity matrix, omitting very low similarities. Batch recommendation can be completed in a map-only phase if the similarity matrix fits into the memory. Otherwise a reduce phase is used. To reduce the algorithm's cost, which is dominated by the 'power users', only a randomly selected part of their interactions is contributing to the recommendation. Sensitivity analysis is given for the effects of the users' interaction reduction, using the MovieLens dataset. Both MovieLens and Flixster datasets are used for measuring the algorithm's accuracy using the MAE metric as well as the RMSE for various values of power users' interaction number. Furthermore, the algorithm is evaluated by means of scalability and runtime on Yahoo!Music dataset.

\begin{table}[htp]
\scriptsize
\begin{center}
\begin {tabular}{|c|c|c|c|c|}
\hline
Ref. & Algorithm & Technologies & Datasets & Metrics\\
\hline
\cite{Ub_Hadoop}  &User-based CF &MapReduce Hadoop & Netflix&Speedup  \\ 
\hline
\cite{5693295}  &Parallel user profiling &MapReduce Hadoop &Del.icio.us & Running time \\ 
\hline
\cite{Davidson:2010:YVR:1864708.1864770}  &Distributed item-based & MapReduce& live trafic& CTR (click through rate)\\ 
  &YouTube's &BigTable &(self collected) &long CTR \\ 
  & Recommender& & &Session length \\ 
  & System& & & Recommendation coverage\\ 
&&&&Time until first long watch\\
\hline
\cite{CF_Hadoop} &Item-based CF &MapReduce Hadoop &MovieLens& Isoefficiency,\\
  & & & &Speedup  \\ 
\hline
\cite{multicore_cf} &CF Library (GraphLab) & GraphLab& Yahoo!Music&  RMSE, Speedup\\ 
  & item-KNN, & & &  \\ 
  &time-KNN & & &  \\ 
\hline
\cite{chen}  &User-based clustering& Hadoop&MovieLens &MAE,  \\ 
  &weighted Slope One (CWSO)&Weka & & Acccuracy \\ 
\hline
\cite{Schelter:2012:SSN:2365952.2365984}  &Pairwise item comparison &MapReduce Hadoop &MovieLens &  MAE, RMSE\\ 
  &and top-N recommendation & &Flixter & Speedup,  \\ 
  & & & Yahoo! music& Runtime \\ 
\hline
\end{tabular}
\caption{Memory-based Implementations on Frameworks}
\label{table:MemFrameworks}
\end{center}
\end{table}

\subsection{Model-based Implementations}


A parallel version of the LDA (Latent Dirichlet Allocation) algorithm is presented in \cite{Chen:2009:CFO:1526709.1526801}. LDA's parallelization is accomplished with the MPI library and MapReduce. Subsets of the users and their ratings are divided among tha available machines. Communication and synchronization among the processes is accomplished with MPI, while with MapReduce Map and Reduce functions are defined and disk I/O operations are performed. A detailed description of the MPI based PLDA algorithm and a version on MapReduce are given in \cite{Wang:2009:PPL:1574036.1574062}. The MPI implementation is publicly available, fact that facilitates experimental reproducibility.
 
The only implementation that utilizes the pervasive DataRush library \cite{Datarush} develops a parallel implementation of the  Bregman co-clustering algorithm \cite{co_cluster_netflix}. Both co-clustering training and prediction algorithms are implemented by a dataflow graph. The pervasive DataRush library is used to construct and execute the dataflow graphs. The number of the used cores influences the number of data partitions that will be processed. An evaluation is provided and a few optimizations are proposed, such as the use of JOMP or adjusting the distance computations according to a technique more adequate for sparse data.

A parallel SGD algorithm for MapReduce is described in \cite{NIPS2010_1162}. A method is presented, where stochastic gradient descent runs in parallel on different computers and their results are aggregated. The only communication needed between the computers is during the results collection, thus, only one MapReduce phase is needed. RMSE is the evaluation metric used and experiments run on a dataset formed by an email system.

The SGD algorithm is also approached in \cite{Gemulla}. A stratified variant of SGD is developed and adjusted, in order to obtain the distributed algorithm DSGD. The input data is distributed over the nodes at the beginning of the execution, while smaller matrices are transmitted during the rest of the execution. Each node creates a local training sequence from the data that receives. During each iteration a step size and a stratum is chosen. Then, SGD runs on the training points in such a way that the whole training set is finally covered. For the experiments two clusters are used. A cluster for the in-memory implementation, which is based on R and C and consists of 32 cores, and a Hadoop cluster consisted of 320 cores. The Netflix dataset is used, and speedup and the elapsed wall-clock time are measured.

An extension of the above SGD algorithm is presented in \cite{pcf_streaming_data}. This approach is designed to operate on streaming data and is implemented on a cluster composed of 10 machines, using the Hadoop Map-Reduce and the Storm framework. The master node assigns dynamically data chunks to workers, taking care to avoid the need of simultaneous update of the same rows or columns. To compute a stratum, the input to the Map phase is the ratings matrix and the $U$ and $M$ matrices. If the rating belongs to the current stratum, the mapper outputs the key-value pairs that correspond the stratum blocks to the ratings that they contain. The reducers receive the information that belongs to a stratum block and SGD runs on them. The MovieLens dataset is used and the results are presented on plots of the total elapsed time versus RMSE and of the number of iterations versus RMSE.

An open source collaborative filtering library is implemented in \cite{multicore_cf}, using the GraphLab parallel machine learning framework. The implemented algorithms are ALS, Wals, BPTF, SGD, SVD++, Item-kNN, time-kNN, time-SGD, time-SVD++, MFITR and time-MFITR. Although a few memory-based algorithms are implemented, emphasis is given to the matrix factorization algorithms. Experiments are conducted on a cluster composed of 32 cores and on the BlackLight supercomputer \cite{blacklight} (4096 cores). The RMSE metric is measured on the validation dataset and the speedup is calculated on BlackLight. The Yahoo! Music dataset is used.

In \cite{Kanagal:2012:SRS:2336664.2336669} is developed a parallel multi-core implementation of the taxonomy-aware latent factor model (TF), implemented in C++. The BOOST library is also used. The SGD algorithm is approached by a multithreaded implementation. Using Hadoop, a different part of the set of users is assigned to each node. As a dataset, a log of user online transactions is used. The AUC metric and the average mean rank of the users are used to compare the proposed model with the basic latent factor model. Also, absolute wall-clock times and speedup are measured on a 12 core machine.

In \cite{Schelter:2013:DMF:2507157.2507195} is parallelized the ALS algorithm on MapReduce using a parallel broadcast-join. Each machine has a local part of the matrix $A$ that contains the user's interactions over the set of items. Furthermore, the smaller of the user $U$ or item $M$ feature matrices is replicated to all the available machines. A map phase joins the local part of $A$ and the replicated copy of the feature matrix and recomputes the other feature matrix. The experiments are realized using three datasets. Netflix, Yahoo!Music and Bigflix, which is a synthetic dataset constructed from Netflix dataset. The average runtime for a recomputation of the feature matrix is measured.

In \cite{Tang:2013:SMF:2487788.2487801} a two-stage matrix factorization is proposed. First runs the truncated SVD algorithm on a MapReduce cluster. Then, the ALS algorithm is applied, starting with the matrix that has been received as a result from the truncated SVD instead of using a random matrix $Q$. With one Map-Reduce step the matrix $P$ is calculated. To evaluate this approach two metrics are used. MAP (Mean Average Precision) and NDCG (Normalized Discounted Cumulative Gain). Unfortunately, these metrics are not used in other similar experiments, and no information is given on whether the data that is collected from the Walmart.com site can be publicly available.



\begin{table}[htp]
\scriptsize
\begin{center}
\begin {tabular}{|c|c|c|c|c|}
\hline
Ref. & Algorithm & Technologies & Datasets & Metrics\\
\hline
\cite{Chen:2009:CFO:1526709.1526801}  &Parallel LDA &MPI & Orkut &Scalability, Speedup,  \\ 
 & &  MapReduce & & Running time \\ 
\hline
\cite{Wang:2009:PPL:1574036.1574062}  &PLDA &MPI &Wikipedia &Speedup  \\
 & &  MapReduce&  A forum dataset & Computation time\\
  & & & & Communication time \\ 
   & & & & Running time \\   
\hline
\cite{co_cluster_netflix} &Co-clustering &Pervasive &Netflix &RMSE,\\
 &Dataflow  & DataRush& &Speedup \\ 
  &Bregman &Library & & Prediction/training time\\ 
\hline
\cite{NIPS2010_1162}&SGD & MapReduce& e-mail system& RMSE\\ 
\hline
 \cite{Gemulla}  &Distributed  & R and C,& Netflix&Speedup,  \\ 
  &Stratified  DSGD &Hadoop & & Elapsed wall-clock time \\ 
\hline
  \cite{pcf_streaming_data}& Distributed SGD&MapReduce Hadoop, &MovieLens &Total elapsed time vs RMSE,  \\
  & (Streaming data)& Storm& &Number of iterations vs RMSE  \\  
\hline
 \cite{multicore_cf} &CF Library (GraphLab) & GraphLab& Yahoo!Music&  RMSE, Speedup\\ 
  &ALS, Wals, BPTF, SGD, & & &  \\ 
  & SVD++, time-SGD, & & &  \\ 
  &time-SVD++,MFITR, & & &  \\ 
  &time-MFITR & & &  \\ 
\hline
\cite{Kanagal:2012:SRS:2336664.2336669}  &Multi-core  &C++ & A log of user& AUC,Speedup \\ 
  & (TF) taxonomy-aware&BOOST library & online&Absolute wall-clock time \\ 
  &Latent Factor Model  & Hadoop & transactions &Average mean  \\ 
  & (SGD) & &  &rank of users \\ 
\hline
 \cite{Schelter:2013:DMF:2507157.2507195} & ALS &MapReduce Hadoop & Netflix & Average runtime \\ 
  & & JBlas & Yahoo!Music& per recomputation \\ 
 &  & & Bigflix (synthetic)&  \\ 
\hline
\cite{Tang:2013:SMF:2487788.2487801} & Truncated SVD, & MapReduce & Collected from  & MAP, NDCG \\
  &  ALS &  &  Walmart.com & \\
\hline
\end{tabular}
\caption{Model-based Implementations on Frameworks}
\label{table:ModFrameworks}
\end{center}
\end{table}

\section{Heterogeneous Implementations}
\label{heterogeneous}

A few hybrid collaborative filtering implementations have been recently developed on both shared and distributed memory systems. All of them have been implemented with MPI and OpenMP or Pthreads. To the remaining of this section they will be described starting with the oldest and proceeding to the most recent implementation. In table \ref{table:Hybridcategories} can be seen a list of these approaches, and the datasets used to each implementation as long as the metrics that are considered can be seen in table \ref{table:hybridmd}.

A distributed model-based algorithm based on co-clustering is presented in \cite{IBM_India_11}. The algorithm partitions row and column clusters to the nodes, which are further partitioned to each node's threads. Iterations are executed until reaching the desired RMSE convergence. One thread on each node, apart from contributing to the computations, takes over the necessary communication to collect the results of the computations assumed by the remaining threads. Netflix Prize dataset is used on a 1024-node Blue Gene/P architecture. Training and prediction time are measured as long as the RMSE metric, and a detailed scalability analysis is also presented. 

Other variations on the distributed co-clustering based collaborative filtering algorithm are presented in \cite{IBM_report}. A distributed flat co-clustering algorithm is implemented using MPI and a flat hybrid algorithm is developed using MPI and OpenMP. Hierarchical co-clustering algorithms are also developed. The algorithms are evaluated on the Blue gene/P architecture and the datasets and metrics used can be seen in table \ref{table:hybridmd}.

In \cite{karydi_margaritis_2} a hybrid version of the Slope One algorithm is presented and compared to the multithreaded version which is described in \cite{karydi_margaritis_1}. Parts of the ratings matrix are distributed over the system's nodes. The master-workers model is followed. The master node assumes the data partitioning and distribution, while the worker nodes are devoted to the computations. Finally all the worker's results are gathered to the master node where the predictions are made. This implementation is evaluated on an heterogeneous cluster composed of 30 cores and a homogeneous cluster composed of 24 cores. The MovieLens dataset is used for the performance and scalability evaluation and total elapsed time, speedup, number of predictions per second and prediction time per rating are measured.

A semi-sparse algorithm, which aims in accelerating the common memory-based collaborative filtering algorithms is proposed in \cite{GuanLG12}. Three optimizing methods are applied. First, a semi-sparse algorithm which denses locally the selected sparse vectors is used to speed up the similarity computations. On a multicore architecture, threads are wrapped into a thread-pool and a reduce vector is used to diminish the use of locks. Moreover, to reduce the communication overhead among different nodes, a shared zip file that contains the sparse rating matrix is read. Experiments are conducted on three different multicore systems and on a cluster of 8 nodes.

\begin{table}[htp]
\scriptsize
\begin{center}
\begin {tabular}{|c|c|c|c|}
\hline
\normalsize{Reference} & \normalsize{Year} & \normalsize{Category} & \normalsize{Description}\\
\hline
\cite{IBM_India_11} & 2011 & MODEL & Co-clustering\\
\hline
\cite{IBM_report} & 2011 & MODEL & Co-clustering\\
\hline
\cite{karydi_margaritis_2} & 2012	& MEMORY	& Slope One\\
\hline
\cite{GuanLG12}	& 2012	& MEMORY & Semi-sparse Multilayer Optimization on Item-based	\\
\hline
\end{tabular}
\caption{List of Heterogeneous Implementations}
\label{table:Hybridcategories}
\end{center}
\end{table}

\begin{table}[htp]
\scriptsize
\begin{center}
\begin {tabular}{|c|c|c|c|c|}
\hline
Ref. & Algorithm & Technologies & Datasets & Metrics\\
\hline
 \cite{IBM_India_11} & Distributed &MPI, OpenMP &Netflix &RMSE, Scalability \\ 
  &Co-clustering & & &Training time \\ 
  & & & &Prediction  time per rating \\ 
\hline
\cite{IBM_report} & Distributed &MPI, OpenMP &Netflix & (Weak, strong, data )Scalability \\ 
  &Co-clustering & & Yahoo KDD cup &RMSE\\ 
    &variations& & &\\ 
\hline
\cite{karydi_margaritis_2}   & Slope One& MPI, OpenMP& MovieLens&Scalability, Speedup \\ 
  & & & & Total elapsed time \\ 
  & & & &Prediction per second \\ 
  & & & &Prediction  time per rating \\ 
\hline
 \cite{GuanLG12}&Semi-sparse &MPI & MovieLens& Speedup\\ 
  &Multi-layer optimization &Pthreads & Netflix& Elapsed CPU time \\ 
   &(Item-based) & & & \\ 
\hline
\end{tabular}
\caption{Heterogeneous Implementations}
\label{table:hybridmd}
\end{center}
\end{table}

\section{Discussion and Conclusions }
\label{conclusions}

Since research papers concerning recommender systems are published in a variety of journals and conferences that focus on different disciplines \cite{Park:2012:LRC:2181339.2181690}, it is not easy to ensure that all the existing implementations are considered in this survey. Great effort has been made to include as many as possible. In any case, no change to the conclusions that have arisen from this work is expected if a few more works appear.

The classification of all the implementations that are discussed to the above sections is summarized in table \ref{table:AllImplementations}. An initial observation is that regardless the parallel or distributed method used, less hybrid implementations exist than memory or model-based. Hence, more hybrid algorithms could be developed that would benefit from both categories' advantages. Another fact worth noticing is that no memory-based implementations are developed on distributed-memory systems and only one on a shared-memory environment. This may be due to the high communication cost that is needed when the whole dataset is used. However, since memory-based collaborative filtering algorithms also deliver good results, they should not be left aside.  

\begin{table}[htp]
\scriptsize
\begin{center}
\begin{tabular}{|c||c|c|c|}
\hline
 & \multicolumn{3}{c|}{\textbf{Collaborative Filtering}}\\
\cline{2-4}
 & \textbf{Memory-based} & \textbf{Model-based} & \textbf{Hybrid}\\
\hline\hline
				&	&	&	\\
	&\cite{Tveit:2001:PBR:381461.381466},
\cite{Han2004203}, \cite{Han2004},
\cite{MillerKR04},	&\cite{Harth},
\cite{Canny:2002:CFP:829514.830525},
	&	\cite{Olsson98decentralisedsocial}, \cite{Link05distributedrecommender}, \cite{Castagnos:2006:CUC:1567016.1567150},
\\
\textbf{Distributed}				&\cite{Berkovsky05collaborativefiltering},
\cite{Wang:2006:DCF:1141277.1141522},
\cite{Berkovsky06hierarchicalneighborhood},
\cite{Xie20071349},
\cite{Berkovsky:2007:EPP:1297231.1297234},	&\cite{Isaacman:2011:DRP:2043932.2043948}, 
\cite{abs-1109-3318},
	& 	\cite{AwerbuchPPT05}, \cite{kumar}, \cite{Ali:2004:TMS:1014052.1014097}	\\
				&\cite{Berkovsky:2007:DCF:1297231.1297238},
\cite{Castagnos:2007:PCD:1763653.1763695},
 \cite{Ruffo:2009:PRS:1462159.1462163}, \cite{Ahn:2010:TFD:1913793.1914138}	&	&  	\\
				&	&	&  	\\
\hline\hline
\textbf{Parallel}&	&	&	\\
	&	&	&  	\\
Distributed Memory	&	&\cite{Par_cf_netflix}, \cite{mfrm},
\cite{dmcomp},
	&	\cite{Chen:2008:CCF:1401890.1401909}\\
		&	&\cite{co_cluster_2005}, \cite{kwon},  \cite{Liu:2011:PPL:1961189.1961198},
\cite{6413871}	&  	\\	
		&	&	&  	\\	
		&	&	&  	\\			
Shared Memory	&\cite{karydi_margaritis_1}	&\cite{IBM_India_10},\cite{recht},\cite{RechtRWN11},\cite{LOU1},\cite{mfrm},\cite{Zhuang:2013:FPS:2507157.2507164}	&	\\
	&	&	&  	\\
			&	&	&  	\\	
		
GPU		&\cite{Kato:2010:SKN:1844765.1845125},
 \cite{Li:2011:SNT:1998076.1998131},
\cite{6337114},
\cite{6384989}	& \cite{Bondhugula},
\cite{5161058},
\cite{katosvdgpu},
\cite{6064611},
\cite{ChuaGpuR},	&  	\\
	&	&\cite{Cai:2012:GRB:2403514.2403537}, 
\cite{Zastrau:2012:SGD:2406479.2406497},
\cite{Foster:2011:GAS:2351958.2352024},
\cite{GPU14}	&  	\\	
	&	&	&  	\\
\hline\hline
&	&	&	\\
\textbf{Platform-based}	&\cite{Ub_Hadoop},
\cite{5693295},
\cite{Davidson:2010:YVR:1864708.1864770},
\cite{CF_Hadoop},	&\cite{Chen:2009:CFO:1526709.1526801},
\cite{Wang:2009:PPL:1574036.1574062},
\cite{co_cluster_netflix},
\cite{NIPS2010_1162},
\cite{Gemulla},	& \cite{Das:2007:GNP:1242572.1242610}	\\
	&\cite{multicore_cf},
\cite{chen},
\cite{Schelter:2012:SSN:2365952.2365984}	&\cite{pcf_streaming_data},
\cite{Kanagal:2012:SRS:2336664.2336669},
\cite{Schelter:2013:DMF:2507157.2507195},
\cite{Tang:2013:SMF:2487788.2487801}, \cite{multicore_cf}
	&  	\\
&	&	&	\\
\hline\hline
&	&	&	\\
\textbf{Heterogeneous}	& \cite{karydi_margaritis_2},
\cite{GuanLG12}	&\cite{IBM_India_11},
\cite{IBM_report}	&	\\
	&	&	&  	\\
\hline
\end{tabular}
\caption{Classification of all the Implementations}
\label{table:AllImplementations}
\end{center}
\end{table}


Table \ref{table:Distributedcategories} lists by chronological order all the distributed collaborative filtering implementations discussed in section \ref{surveydistributed}. An initial preference to the memory-based techniques is observed. However, during the most recent years the interest seems to turn to model-based and hybrid approaches. This probably occurs because the dimensionality reduction techniques are more suitable to cope with the all-increasing amount of data to be processed. Thus, the model-based approaches seem to be more promising to deliver results faster than memory based approaches. 

Among the memory-based algorithms, traditional user and item-based algorithms are deployed more often than the top-N approaches. The majority of the distributed memory-based collaborative filtering approaches employ the MAE metric to measure the recommendations' accuracy. Other metrics are being used less, such as recall, coverage and precision. However, none of the experiments includes speedup analysis and computation or communication time are scarcely considered. Emphasis is given to privacy issues by distributing  parts of the users' information to the available peers. Occasionally, the peer-to-peer architecture is simulated by multithreaded applications, though no preference to any specific  technology is shown. The MovieLens and the EachMovie datasets are preferred on the larger part of the experiments.

The model-based algorithms that are developed on distributed systems are not enough to offer sufficient conclusions. However, it is noticeable that none of the implementations employs  clustering techniques and the dimensionality reduction techniques seem to attract more interest. Though no preference is shown to any specific metrics, in these approaches, except from the accuracy metrics are also measured factors such as the time needed for a recommendation or the algorithm's convergence. Furthermore, in \cite{Harth} and \cite{abs-1109-3318} are proposed models that use disjoint datasets. This fact can improve a method's security, since data is not gathered to any specific peer.

Table \ref{table:DistrHybrid} shows the hybrid distributed collaborating filtering approaches. As can be seen, information on the technologies and datasets used is incomplete and no common framework exists on the performance evaluation. Some of the proposed methods are mathematical simulations and are not implemented. Also, the small number of hybrid distributed approaches reveals a gap that needs to be filled. Investigating the performance of other hybrid implementations could be proved useful.

In all parallel implementations a clear preference to the model-based algorithms is observed. As can be seen in table \ref{table:Parallelcategories}, the majority of the algorithms that are implemented on distributed memory systems are model-based. Only one hybrid approach exists, and none of the approaches implements memory-based algorithms. A possible explanation for this fact is that memory-based algorithms need to process the whole dataset, thus inter-node communication on the cluster would be prohibitively expensive.

The algorithms that are more often implemented on distributed-memory systems are ALS, SGD and co-clustering methods. MPI is used for communication among the system's nodes in all the implementations. To these approaches, speedup is the metric that is most often used for evaluation. The Netflix dataset is used to almost all the implementations, followed by MovieLens and the KDD Cup 2011 dataset.
 
The shared-memory collaborative filtering implementations are listed according to their publication year in table \ref{table:Multithreadedcategories}. Though very few shared-memory approaches have been found to be able to draw significant conclusions, one interesting fact is that all the approaches are very recent. A preference to the model-based approaches is shown, without indicating any inclination to a specific algorithm. The Netflix, MovieLens and Yahoo!Music datasets are used to conduct experiments, observing that to the most recent implementations all the three datasets are used to provide more accurate explanations on the results.

Time related measurements seem to be more important in the shared-memory implementations than in the distributed-memory implementations, having speedup and scalability analysed in almost all the implementations. The RMSE metric is also taken under consideration by the majority of the implementations, while in none of them are conducted experiments using the MAE metric. Furthermore, it is important to observe that none of the shared-memory implementations combines model and memory-based algorithms. 
 
All the implementations that are developed using GPUs are built on CUDA. A preference to the model-based algorithms is also shown to the implementations that take advantage of GPU accelerators. Most of the memory-based applications parallelize the user-based algorithm and a few are dealing with the item-based and neighbourhood-based algorithms. However, the memory-based implementations are too few to allow for sufficient conclusions. The datasets preferred are the Flixster, the MovieLens and the Bookcrossing.

Regarding the metrics used, a preference is noticed for the measurement of the total execution time and the speedup over the sequential implementations. RMSE is concerning less the researchers since no attention is given to prove the selected algorithms' efficiency. They are rather concerned to compare the CPU and GPU execution times. For the first time is noticed a focus on the power and energy consumption of the implementations. Although this metric is approached only by one implementation, other works are also expected to concern such issues in the future.

Among the model-based collaborative filtering algorithms, the one that has been consistently selected for parallelization on GPUs is the SVD. Other algorithms, such as SGD, co-clustering and usage of restricted Boltzmann machines on collaborative filtering have been also implemented, though not in such an extend. The majority of the model-based algorithms that have been implemented using CUDA employ libraries, such as CUBLAS or CULA, to handle more efficiently the various algebraic problems that they encounter.

It is interesting that many of the model-based approaches on GPU select random datasets for the experiments. This fact except form negatively affect the experimental reproducibility, also complicates the comparison of the results to those of other implementations. Except from the randomly produced datasets, the Netflix dataset is the most used. The metrics that are mostly preferred are speedup and execution time. Measuring computation and communication time, as long as RMSE, very scarcely occurs. When using real big datasets on GPUs the problem of high data transfer time between CPU and GPU occurs and can significantly affect the overall performance. Fortunately,  major companies of the field have recently announced developments on new technologies that can face this challenge via unified memory \cite{NvidiaUM}, \cite{AMDUM}. Consequently, GPUs are expected to be used more extensively for the development of applications that will take advantage of the information provided by real big datasets.

It is interesting that among the platform-based implementations only one hybrid implementation combining both model and memory-based techniques is observed. Also, there is no definite trend in favour of one of the two categories. Both model and memory-based algorithms have been chosen for implementation on frameworks. The algorithms that are more often employed among the model-based implementations on frameworks are LDA, SGD and SVD and among the memory-based algorithms, the user-based and item-based collaborative filtering.

A variety of datasets is used to evaluate the discussed approaches. Although many approaches are evaluated on unusual datasets, the dominating datasets are Netflix, MovieLens and Yahoo!Music. The majority of the applications are implemented on the Hadoop MapReduce framework, but especially to the model-based implementations, some approaches that combine Hadoop with other parallel computing libraries, such as MPI or pervasive DataRush have been developed. Also, many algorithms have been implemented on GraphLab.

Concerning the metrics most commonly used, a preference is noticed to RMSE, MAE, running time and speedup. A fact that proves the comparison of all the implementations a difficult task is that they have been executed on systems that significantly differ on the number of used cores.

Furthermore, the heterogeneous implementations that combine several parallelization techniques are very few and only distributed-memory approaches are combined with shared-memory models. More heterogeneous implementations could be developed, combining various parallelization techniques.

Unfortunately, the results of the heterogeneous implementations cannot compared to each other, not only because of the use of different datasets, but also because of the use of different cluster architectures with significant difference in the number of nodes. No preference is given neither to model-based nor to memory-based algorithms. In all these implementations the communication among the cluster nodes is accomplished with MPI, while OpenMP or Pthreads are used for shared memory parallelization. The dominating datasets are the Netflix and the MovieLens datasets. To these implementations priority is given in measuring their scalability and speedup, while measuring the algorithm's accuracy by means of the RMSE metric is of less interest.

Table \ref{table:ImplAlgorithms} presents a list of the algorithms of each category that have been implemented using a parallel or distributed computing technique and each algorithm's implementations. In table \ref{table:AllAlgorithms} more information can be seen on the parallelization techniques that are used on each algorithm's implementations. Among the memory-based algorithms the user-based algorithm is more implemented, followed by the item-based algorithm. The most frequently implemented model-based algorithms are SVD, SGD, ALS and co-clustering models. Among the implementations of hybrid algorithms is not distinguished any specific scheme.

The present work verifies the fact that the field of parallel and distributed collaborative filtering is active and evolves quickly. Great effort has been made to comprise as many implementations as possible, that have been published in scientific journals or conferences before the end of 2013. Furthermore, a category of papers which make small use of recommender systems while their main focus is on neural networks or other artificial intelligence techniques has been omitted from the present work.

Recently many parallel and distributed collaborative filtering approaches have been developed, especially employing GPUs or taking advantage of various platforms. Yet, more research needs to be conducted in order to exploit the benefits of parallel and distributed computing techniques  and improve the collaborative filtering algorithms in such way as to handle more efficiently the huge amounts of data that are available.

It would be interesting to apply a multi-level heterogeneous method, using many machines to efficiently handle big data and subsequently combine a variety of techniques according to the addressed problem. In recent years, although a variety of parallel and distributed techniques is applied, a preference is noticed to the usage of graphics accelerators and frameworks. Thus, the usage of an adequate framework in combination with MPI and GPU accelerators would be intriguing. An aspect that is determinant for the selection of a technique is the nature of the available data. If it is hard to collect all the data in one machine, then distributed methods should be preferred, while clusters or methods based on shared-memory environments are more adequate when data is easily assembled in one place.

Moreover, the strategy mentioned above could be applied over a distributed recommender systems, where data could be either divided to the system's nodes or available to all system's nodes. Each node could apply different algorithms over different technologies or platforms. Finally, a node could be assigned the task of collecting and elaborating the results from all the system's nodes and providing the final recommendations. Such a system would have the advantages of preserving privacy, multifaceted data processing over various algorithms and simultaneous usage of different technologies, and would favour the recommendation of items of different nature as long as the usage of data of different structure.

Briefly summarizing the findings of the research work discussed in this article, the preference of the research community to the development of model-based collaborative filtering algorithms is clear. Memory-based and especially hybrid algorithms are implemented less. Still, the development of hybrid algorithms seems promising, since advantage could be taken of both method's benefits.

In recent years, a trend to the usage of frameworks and GPU accelerators has been noticed, having MPI-based and shared-memory techniques in second place. The usage of frameworks is anticipated to be more flexible in the future and combined with other techniques.

As far as the evaluation of the implementations is concerned,  initially algorithmic accuracy was the main interest, which was measured by MAE metric. Lately, the interest has turned towards scalability analysis and the achievement of fewer execution time. A few approaches are tested on self-collectedMoreover, the strategy mentioned above could be applied over a distributed recommender systems, where data could be either divided to the system's nodes or available to all system's nodes. Each node could apply different algorithms over different technologies or platforms. Finally, a node could be assigned the task of collecting and elaborating the results from all the system's nodes and providing the final recommendations. Such a system would have the advantages of preserving privacy, multifaceted data processing over various algorithms and simultaneous usage of different technologies, and would favor the recommendation of items of different nature as long as the usage of data of different structure. data, which are not publicly available for further experiments. However, the majority of the implementations is tested on the well known Netflix, MovieLens and Yahoo!Music datasets.

As a conclusion, new technologies are continuously contributing to the development of parallel and distributed collaborative filtering algorithms. There is no specific pattern to be followed, since the selection of the adequate technology is highly related to the nature of the available data, the characteristics of the algorithms and the available hardware. The work discussed in this article is expected to provide a useful starting basis, to offer helpful directions for both the selection of technologies and algorithms, and to trigger inspiration for the development of more sophisticated approaches.

\begin{table}[htp]
\tiny
\begin{center}
\begin{tabular}{|c|c|}
\hline
\textbf{Algorithm} & \textbf{References} \\
\hline
\textbf{Memory-based}&  \\
&  \\
User-based & \cite{Tveit:2001:PBR:381461.381466}, \cite{Han2004203}, \cite{Han2004}, \cite{Berkovsky05collaborativefiltering}, \cite{Berkovsky06hierarchicalneighborhood}, \cite{Xie20071349}, \cite{Berkovsky:2007:EPP:1297231.1297234}, \cite{Castagnos:2007:PCD:1763653.1763695},  \cite{Ruffo:2009:PRS:1462159.1462163}, \cite{Ahn:2010:TFD:1913793.1914138}, \cite{6337114}, \cite{6384989}, \cite{Ub_Hadoop}, \cite{5693295}, \cite{chen} \\
&  \\
Item-based & \cite{MillerKR04}, \cite{Berkovsky:2007:DCF:1297231.1297238}, \cite{6337114}, \cite{Davidson:2010:YVR:1864708.1864770}, \cite{CF_Hadoop}, \cite{GuanLG12} \\
&  \\
User-based top-N & \cite{Li:2011:SNT:1998076.1998131} \\
&  \\
Item-based top-N &\cite{Wang:2006:DCF:1141277.1141522}, \cite{Schelter:2012:SSN:2365952.2365984}\\
&  \\
Slope One & \cite{karydi_margaritis_1}, \cite{chen}, \cite{karydi_margaritis_2}\\
& \\
K-nearest neighbour & \cite{Kato:2010:SKN:1844765.1845125}, \cite{multicore_cf} \\
& \\
\hline
\textbf{Model-based}&  \\
&  \\
SVD & \cite{Canny:2002:CFP:829514.830525}, \cite{Bondhugula}, \cite{5161058}, \cite{katosvdgpu}, \cite{Tang:2013:SMF:2487788.2487801} \\
&	\\
SVD++ & \cite{multicore_cf} \\
& \\
Aproximate SVD & \cite{ChuaGpuR}, \cite{Foster:2011:GAS:2351958.2352024}\\
&  \\
SGD & \cite{Isaacman:2011:DRP:2043932.2043948}, \cite{RechtRWN11}, \cite{recht}, \cite{LOU1}, \cite{Zastrau:2012:SGD:2406479.2406497}, \cite{multicore_cf}, \cite{NIPS2010_1162}, \cite{Gemulla}, \cite{pcf_streaming_data} \\
&  \\
ALS & \cite{multicore_cf}, \cite{Schelter:2013:DMF:2507157.2507195}, \cite{Tang:2013:SMF:2487788.2487801}, \cite{dmcomp}\\
& \\
ALS-WR &\cite{Par_cf_netflix}  \\
&  \\
CCD++ & \cite{mfrm}  \\
&  \\
ASGD &  \cite{dmcomp} \\
& \\
DSGD++  &\cite{dmcomp} \\
& \\
FPSGD & \cite{Zhuang:2013:FPS:2507157.2507164} \\
& \\
LDA & \cite{Chen:2009:CFO:1526709.1526801}\\
& \\
PLDA & \cite{Wang:2009:PPL:1574036.1574062}\\
& \\
PLDA+ & \cite{Liu:2011:PPL:1961189.1961198}\\
& \\
Bregman Co-clustering & \cite{co_cluster_2005}, \cite{kwon}, \cite{co_cluster_netflix} \\
 & \\
 Co-clustering & \cite{6413871}, \cite{6064611}, \cite{IBM_India_11}, \cite{IBM_report} \\
  & \\
User profiling probabilistic model & \cite{abs-1109-3318}  \\
&  \\
Association rules & \cite{Harth}  \\
& \\
Concept Decomposition & \cite{IBM_India_10} \\
& \\
RBM-CF & \cite{Cai:2012:GRB:2403514.2403537}, \cite{GPU14} \\
&	\\
Taxonomy-aware & \cite{Kanagal:2012:SRS:2336664.2336669} \\
Latent factor & \\
&	\\
\hline
\textbf{Hybrid}&  \\
&  \\
Content-based, Item-based  & \cite{Olsson98decentralisedsocial}  \\
and Social filtering & \\
&  \\
Content-based and  & \cite{Link05distributedrecommender}\\
neighbourhood-based& \\
&  \\
Hierarchical clustering & \cite{Castagnos:2006:CUC:1567016.1567150} \\
and user-based & \\
& \\
Random Product or& \cite{AwerbuchPPT05} \\
User Probation & \\
& \\
CAPSSR & \cite{kumar}\\
& \\
Item-based and Bayesian & \cite{Ali:2004:TMS:1014052.1014097} \\
 Content-based filtering &  \\
& \\
Combinational CF & \cite{Chen:2008:CCF:1401890.1401909} \\
& \\
MinHash and PLSI & \cite{Das:2007:GNP:1242572.1242610} \\
CLustering & \\
& \\
\hline
\end{tabular}
\caption{Implemented Algorithms}
\label{table:ImplAlgorithms}
\end{center}
\end{table}

\begin{table}[htp]
\tiny
\begin{center}
\begin{tabular}{|c|c|c|c|c|c|c|}
\hline
\multirow{3}{*}{\backslashbox{\textbf{Algorithm}}{\textbf{Parallelization}\\ \textbf{Technique}}  }& \textbf{Distributed }&  \multicolumn{3}{c|}{\textbf{Parallel}} & \textbf{Platform} & \textbf{Heterogeneous} \\
\cline{3-5} 
& & \textbf{Distributed} & \textbf{Shared} &  \textbf{GPU} & \textbf{based} & \\
& & \textbf{memory} & \textbf{memory} &   & & \\
\hline

\textbf{Memory-based}& & & & & & \\
& & & & & & \\
User-based & \cite{Tveit:2001:PBR:381461.381466}, \cite{Han2004203}, \cite{Han2004}  & & &\cite{6337114}, \cite{6384989} & \cite{Ub_Hadoop}, \cite{5693295} & \\
&\cite{Berkovsky05collaborativefiltering}, \cite{Berkovsky06hierarchicalneighborhood}, \cite{Xie20071349} & & & &\cite{chen} & \\
& \cite{Berkovsky:2007:EPP:1297231.1297234}, \cite{Castagnos:2007:PCD:1763653.1763695},  \cite{Ruffo:2009:PRS:1462159.1462163} & & & & & \\
& \cite{Ahn:2010:TFD:1913793.1914138} & & & & & \\
& & & & & & \\
Item-based & \cite{MillerKR04}, \cite{Berkovsky:2007:DCF:1297231.1297238} & & &\cite{6337114} & \cite{Davidson:2010:YVR:1864708.1864770}, \cite{CF_Hadoop} & \cite{GuanLG12} \\
& & & & & & \\
User-based top-N & & & & \cite{Li:2011:SNT:1998076.1998131} & & \\
& & & & & & \\
Item-based top-N &\cite{Wang:2006:DCF:1141277.1141522} & & & &\cite{Schelter:2012:SSN:2365952.2365984} & \\
& & & & & & \\
Slope One & & & \cite{karydi_margaritis_1} & &\cite{chen} & \cite{karydi_margaritis_2} \\
& & & & & & \\
K-nearest neighbour & & & & \cite{Kato:2010:SKN:1844765.1845125}& \cite{multicore_cf}& \\
& & & & & & \\
\hline
\textbf{Model-based}& & & & & & \\
& & & & & & \\
SVD & \cite{Canny:2002:CFP:829514.830525} & & &\cite{Bondhugula}, \cite{5161058}, \cite{katosvdgpu} & \cite{Tang:2013:SMF:2487788.2487801}& \\
& & & & & & \\
SVD++ & & & & &\cite{multicore_cf} & \\
& & & & & & \\
Aproximate SVD & & & & \cite{ChuaGpuR}, \cite{Foster:2011:GAS:2351958.2352024} & & \\
& & & & & & \\
SGD & \cite{Isaacman:2011:DRP:2043932.2043948} &  & \cite{recht},\cite{RechtRWN11},\cite{LOU1} &\cite{Zastrau:2012:SGD:2406479.2406497} &\cite{multicore_cf}, \cite{NIPS2010_1162}, \cite{Gemulla}, \cite{pcf_streaming_data} & \\
& & & & & & \\
ALS & &\cite{dmcomp} & & &\cite{multicore_cf}, \cite{Schelter:2013:DMF:2507157.2507195}, \cite{Tang:2013:SMF:2487788.2487801} & \\
& & & & & & \\
ALS-WR & & \cite{Par_cf_netflix}& & & & \\
& & & & & & \\
CCD++ & &\cite{mfrm} &\cite{mfrm} & & & \\
& & & & & & \\
ASGD & & \cite{dmcomp}& & & & \\
& & & & & & \\
DSGD++ & &\cite{dmcomp}& & & & \\
& & & & & & \\
FPSGD & & & \cite{Zhuang:2013:FPS:2507157.2507164}& & & \\
& & & & & & \\
LDA & & & & &\cite{Chen:2009:CFO:1526709.1526801} & \\
& & & & & & \\
PLDA & & & & & \cite{Wang:2009:PPL:1574036.1574062} & \\
& & & & & & \\
PLDA+ &  & \cite{Liu:2011:PPL:1961189.1961198}& & & & \\
& & & & & & \\
Bregman Co-clustering & & \cite{co_cluster_2005}, \cite{kwon} & & & \cite{co_cluster_netflix}& \\
& & & & & & \\
Co-clustering & & \cite{6413871} & &\cite{6064611} & &\cite{IBM_India_11}, \cite{IBM_report} \\
& & & & & & \\
User profiling  & \cite{abs-1109-3318}& & & & & \\
probabilistic model& & & & & & \\
& & & & & & \\
Association rules & \cite{Harth} & & & & & \\
& & & & & & \\
Concept Decomposition & & & \cite{IBM_India_10}& & & \\
& & & & & & \\
RBM-CF & & & &\cite{Cai:2012:GRB:2403514.2403537}, \cite{GPU14} & & \\
& & & & & & \\
Taxonomy-aware& & & & &\cite{Kanagal:2012:SRS:2336664.2336669} & \\
Latent factor& & & & & & \\
& & & & & & \\

\hline
\textbf{Hybrid}& & & & & & \\
& & & & & & \\
Content-based, Item-based  & \cite{Olsson98decentralisedsocial} & & & & & \\
and Social filtering& & & & & & \\
& & & & & & \\
Content-based and  & \cite{Link05distributedrecommender} & & & & & \\
neighbourhood-based& & & & & & \\
& & & & & & \\
Hierarchical clustering & \cite{Castagnos:2006:CUC:1567016.1567150} & & & & & \\
and user-based & & & & & & \\
& & & & & & \\
Random Product or& \cite{AwerbuchPPT05}& & & & & \\
User Probation & & & & & & \\
& & & & & & \\
CAPSSR & \cite{kumar} & & & & & \\
& & & & & & \\
Item-based and Bayesian & \cite{Ali:2004:TMS:1014052.1014097} & & & & & \\
 Content-based filtering& & & & & & \\
& & & & & & \\
Combinational CF & & \cite{Chen:2008:CCF:1401890.1401909}& & & & \\
& & & & & & \\
MinHash and PLSI& & & & &\cite{Das:2007:GNP:1242572.1242610} & \\
clustering& & & & & & \\
& & & & & & \\
\hline
\end{tabular}
\caption{Parallelization Techniques of the Implemented Algorithms}
\label{table:AllAlgorithms}
\end{center}
\end{table}

\cleardoublepage
\bibliography{citations}{}

\begin{thebibliography}{100}

\bibitem{Datarush}
Analytics engine for parallel data processing: Actian datarush.
\newblock
  \url{http://bigdata.pervasive.com/Products/Analytic-Engine-Actian-DataRush.a%
spx}.

\bibitem{bookcross}
The book-crossing dataset.
\newblock \url{http://www.informatik.uni-freiburg.de/~cziegler/BX/}.

\bibitem{Cascading}
Cascading: Big data application development.
\newblock \url{http://www.cascading.org/}.

\bibitem{flixster}
The flixster dataset.
\newblock \url{http://www.cs.sfu.ca/~sja25/personal/datasets/}.

\bibitem{Adomavicius2005}
Gediminas Adomavicius and Alexander Tuzhilin.
\newblock Toward the next generation of recommender systems: A survey of the
  state-of-the-art and possible extensions.
\newblock {\em IEEE Trans. on Knowl. and Data Eng.}, 17(6):734--749, June 2005.

\bibitem{Ahn:2010:TFD:1913793.1914138}
Jae-wook Ahn and Xavier Amatriain.
\newblock Towards fully distributed and privacy-preserving recommendations via
  expert collaborative filtering and restful linked data.
\newblock In {\em Proceedings of the 2010 IEEE/WIC/ACM International Conference
  on Web Intelligence and Intelligent Agent Technology - Volume 01}, WI-IAT
  '10, pages 66--73, Washington, DC, USA, 2010. IEEE Computer Society.

\bibitem{Ali:2004:TMS:1014052.1014097}
Kamal Ali and Wijnand van Stam.
\newblock Tivo: making show recommendations using a distributed collaborative
  filtering architecture.
\newblock In {\em Proceedings of the tenth ACM SIGKDD international conference
  on Knowledge discovery and data mining}, KDD '04, pages 394--401, New York,
  NY, USA, 2004. ACM.

\bibitem{survey5}
Dhoha Almazro, Ghadeer Shahatah, Lamia Albdulkarim, Mona Kherees, Romy
  Martinez, and William Nzoukou.
\newblock A survey paper on recommender systems.
\newblock {\em CoRR}, abs/1006.5278, 2010.

\bibitem{Amatriain:2009:WFC:1571941.1572033}
Xavier Amatriain, Neal Lathia, Josep~M. Pujol, Haewoon Kwak, and Nuria Oliver.
\newblock The wisdom of the few: a collaborative filtering approach based on
  expert opinions from the web.
\newblock In {\em Proceedings of the 32nd international ACM SIGIR conference on
  Research and development in information retrieval}, SIGIR '09, pages
  532--539, New York, NY, USA, 2009. ACM.

\bibitem{IBM_report}
Abhinav~Srivastava Ankur~Narang and Naga Praveen~Kumar Katta.
\newblock High performance distributed co-clustering and collaborative
  filtering.
\newblock
  \url{http://domino.watson.ibm.com/library/Cyberdig.nsf/papers/E9F8290F6B662A%
EC8525795300452695}.
\newblock 2011.

\bibitem{audioscrobbler}
Audioscrobbler.
\newblock \url{http://www.audioscrobbler.com}.

\bibitem{AwerbuchPPT05}
Baruch Awerbuch, Boaz Patt-Shamir, David Peleg, and Mark~R. Tuttle.
\newblock Improved recommendation systems.
\newblock In {\em SODA}, pages 1174--1183, 2005.

\bibitem{Banerjee_Dhillon_Ghosh_Merugu_Modha_2004}
Arindam Banerjee, Inderjit Dhillon, Joydeep Ghosh, Srujana Merugu, and
  Dharmendra~S Modha.
\newblock A generalized maximum entropy approach to bregman co-clustering and
  matrix approximation.
\newblock {\em Proceedings of the 2004 ACM SIGKDD international conference on
  Knowledge discovery and data mining KDD 04}, 8:509, 2004.

\bibitem{Bellogín2013142}
Alejandro Bellog\'{\i}n, Iv\'{a}n Cantador, and Pablo Castells.
\newblock A comparative study of heterogeneous item recommendations in social
  systems.
\newblock {\em Information Sciences}, 221(0):142 -- 169, 2013.

\bibitem{Bellogin:2013:ECS:2414425.2414439}
Alejandro Bellog\'{\i}n, Iv\'{a}n Cantador, Fernando D\'{\i}ez, Pablo Castells,
  and Enrique Chavarriaga.
\newblock An empirical comparison of social, collaborative filtering, and
  hybrid recommenders.
\newblock {\em ACM Trans. Intell. Syst. Technol.}, 4(1):14:1--14:29, February
  2013.

\bibitem{Berkovsky05collaborativefiltering}
Shlomo Berkovsky, Paolo Busetta, Yaniv Eytani, Tsvi Kuflik, and Francesco
  Ricci.
\newblock Collaborative filtering over distributed environment.
\newblock In {\em in proc. of the DASUM Workshop}, 2005.

\bibitem{Berkovsky:2007:EPP:1297231.1297234}
Shlomo Berkovsky, Yaniv Eytani, Tsvi Kuflik, and Francesco Ricci.
\newblock Enhancing privacy and preserving accuracy of a distributed
  collaborative filtering.
\newblock In {\em Proceedings of the 2007 ACM conference on Recommender
  systems}, RecSys '07, pages 9--16, New York, NY, USA, 2007. ACM.

\bibitem{Berkovsky06hierarchicalneighborhood}
Shlomo Berkovsky and Tsvi Kuflik.
\newblock Hierarchical neighborhood topology for privacy enhanced collaborative
  filtering.
\newblock In {\em In proceedings of the PEP06, CHI06 Workshop on
  Privacy-Enhanced Personalization}, pages 6--13, 2006.

\bibitem{Berkovsky:2007:DCF:1297231.1297238}
Shlomo Berkovsky, Tsvi Kuflik, and Francesco Ricci.
\newblock Distributed collaborative filtering with domain specialization.
\newblock In {\em Proceedings of the 2007 ACM conference on Recommender
  systems}, RecSys '07, pages 33--40, New York, NY, USA, 2007. ACM.

\bibitem{blacklight}
Blacklight.
\newblock \url{http://www.psc.edu/machines/sgi/uv/blacklight.php}.

\bibitem{Bobadilla:2013:RSS:2483330.2483573}
J.~Bobadilla, F.~Ortega, A.~Hernando, and A.~Guti{\'e}Rrez.
\newblock Recommender systems survey.
\newblock {\em Knowledge - Based Systems.}, 46:109--132, July 2013.

\bibitem{Burke2002}
Robin Burke.
\newblock Hybrid recommender systems: Survey and experiments.
\newblock {\em User Modeling and User-Adapted Interaction}, 12(4):331--370,
  November 2002.

\bibitem{dmcomp}
R.~Gemulla C.~Teflioudi, F.~Makari.
\newblock Distributed matrix completion.
\newblock {\em in Proceedings of the IEEE International Conference on Data
  Mining (ICDM)}, 2012.

\bibitem{Cai:2012:GRB:2403514.2403537}
Xianggao Cai, Zhanpeng Xu, Guoming Lai, Chengwei Wu, and Xiaola Lin.
\newblock Gpu-accelerated restricted boltzmann machine for collaborative
  filtering.
\newblock In {\em Proceedings of the 12th international conference on
  Algorithms and Architectures for Parallel Processing - Volume Part I},
  ICA3PP'12, pages 303--316, Berlin, Heidelberg, 2012. Springer-Verlag.

\bibitem{GPU14}
Xianggao Cai, Zhanpeng Xu, Guoming Lai, Chengwei Wu, and Xiaola Lin.
\newblock Design and implementation of large scale parallel collaborative
  filtering on multi-core cpu and gpu.
\newblock Submitted to Journal of Parallel and Distributed Computing, January
  2013.

\bibitem{Candillier:2007:CSC:1420326.1420378}
Laurent Candillier, Frank Meyer, and Marc Boull{\'e}.
\newblock Comparing state-of-the-art collaborative filtering systems.
\newblock In {\em Proceedings of the 5th International Conference on Machine
  Learning and Data Mining in Pattern Recognition}, MLDM '07, pages 548--562,
  Berlin, Heidelberg, 2007. Springer-Verlag.

\bibitem{Canny:2002:CFP:829514.830525}
John Canny.
\newblock Collaborative filtering with privacy.
\newblock In {\em Proceedings of the 2002 IEEE Symposium on Security and
  Privacy}, SP '02, pages 45--, Washington, DC, USA, 2002. IEEE Computer
  Society.

\bibitem{Cantador:RecSys2011}
Iv\'{a}n Cantador, Peter Brusilovsky, and Tsvi Kuflik.
\newblock 2nd workshop on information heterogeneity and fusion in recommender
  systems (hetrec 2011).
\newblock In {\em Proceedings of the 5th ACM conference on Recommender
  systems}, RecSys 2011, New York, NY, USA, 2011. ACM.

\bibitem{Castagnos:2006:CUC:1567016.1567150}
Sylvain Castagnos and Anne Boyer.
\newblock A client/server user-based collaborative filtering algorithm: Model
  and implementation.
\newblock In {\em Proceedings of the 2006 conference on ECAI 2006: 17th
  European Conference on Artificial Intelligence August 29 -- September 1,
  2006, Riva del Garda, Italy}, pages 617--621, Amsterdam, The Netherlands, The
  Netherlands, 2006. IOS Press.

\bibitem{Castagnos:2007:PCD:1763653.1763695}
Sylvain Castagnos and Anne Boyer.
\newblock Personalized communities in a distributed recommender system.
\newblock In {\em Proceedings of the 29th European conference on IR research},
  ECIR'07, pages 343--355, Berlin, Heidelberg, 2007. Springer-Verlag.

\bibitem{AMDUM}
AMD~Developer Central.
\newblock huma - the next big thing in processors.
\newblock \url{http://developer.amd.com/}.

\bibitem{Chen:2009:CFO:1526709.1526801}
Wen-Yen Chen, Jon-Chyuan Chu, Junyi Luan, Hongjie Bai, Yi~Wang, and Edward~Y.
  Chang.
\newblock Collaborative filtering for orkut communities: discovery of user
  latent behavior.
\newblock In {\em Proceedings of the 18th international conference on World
  wide web}, WWW '09, pages 681--690, New York, NY, USA, 2009. ACM.

\bibitem{Chen:2008:CCF:1401890.1401909}
Wen-Yen Chen, Dong Zhang, and Edward~Y. Chang.
\newblock Combinational collaborative filtering for personalized community
  recommendation.
\newblock In {\em Proceedings of the 14th ACM SIGKDD international conference
  on Knowledge discovery and data mining}, KDD '08, pages 115--123, New York,
  NY, USA, 2008. ACM.

\bibitem{chen}
X.~Chen and W~Hongfa.
\newblock Clustering weighted slope one for distributed parallel computing.
\newblock {\em Computer Science and Network Technology (ICCSNT)}, 3:1595 --
  1598, 2011.

\bibitem{ChuaGpuR}
Jack Chua.
\newblock Scaling machine learning algorithms across gpu clusters using r.
\newblock 2012.

\bibitem{co_cluster_netflix}
Walker~Matt Daruru~Srivatsava, Marín~Nena and Ghosh Joydeep.
\newblock Pervasive parallelism in data mining : Dataflow solution to
  co-clustering large and sparse netflix data.
\newblock {\em KDD '09 Proceedings of the 15th ACM SIGKDD international
  conference on Knowledge discovery and data mining}, pages 1115--1123, 2009.

\bibitem{Das:2007:GNP:1242572.1242610}
Abhinandan~S. Das, Mayur Datar, Ashutosh Garg, and Shyam Rajaram.
\newblock Google news personalization: scalable online collaborative filtering.
\newblock In {\em Proceedings of the 16th international conference on World
  Wide Web}, WWW '07, pages 271--280, New York, NY, USA, 2007. ACM.

\bibitem{delicious}
The~Delicious dataset.
\newblock \url{http://www.delicious.com}.

\bibitem{jester}
The Jester Collaborative~Filtering Dataset.
\newblock \url{ http://goldberg.berkeley.edu/jester-data/}.

\bibitem{lastfm}
The~Last.fm dataset.
\newblock \url{http://http://www.lastfm.com}.

\bibitem{Davidson:2010:YVR:1864708.1864770}
James Davidson, Benjamin Liebald, Junning Liu, Palash Nandy, Taylor Van~Vleet,
  Ullas Gargi, Sujoy Gupta, Yu~He, Mike Lambert, Blake Livingston, and
  Dasarathi Sampath.
\newblock The youtube video recommendation system.
\newblock In {\em Proceedings of the fourth ACM conference on Recommender
  systems}, RecSys '10, pages 293--296, New York, NY, USA, 2010. ACM.

\bibitem{yahoods}
Gideon Dror, Noam Koenigstein, Yehuda Koren, and Markus Weimer.
\newblock The yahoo! music dataset and kdd-cup '11.
\newblock {\em Journal of Machine Learning Research - Proceedings Track},
  18:8--18, 2012.

\bibitem{cacheda}
Diego~Fern\'{a}ndez Fidel~Cacheda, Victor~Carneiro and Vreixo Formoso.
\newblock Comparison of collaborative filtering algorithms:limitations of
  current techniques and proposals for scalable, high performance recommender
  systems.
\newblock {\em ACM Transactions on the Web}, vol. 05,(No. 1), February 2011.

\bibitem{Foster:2011:GAS:2351958.2352024}
Blake Foster, Sridhar Mahadevan, and Rui Wang.
\newblock A gpu-based approximate svd algorithm.
\newblock In {\em Proceedings of the 9th international conference on Parallel
  Processing and Applied Mathematics - Volume Part I}, PPAM'11, pages 569--578,
  Berlin, Heidelberg, 2012. Springer-Verlag.

\bibitem{Gemulla}
Rainer Gemulla, Erik Nijkamp, Peter~J. Haas, and Yannis Sismanis.
\newblock Large-scale matrix factorization with distributed stochastic gradient
  descent.
\newblock {\em Proceedings of the 17th ACM SIGKDD international conference on
  Knowledge discovery and data mining KDD '11}, pages 69--77, 2011.

\bibitem{co_cluster_2005}
T~George and S~Merugu.
\newblock A scalable collaborative filtering framework based on co-clustering.
\newblock {\em Fifth IEEE International Conference on Data Mining ICDM05},
  pages 625--628, 2005.

\bibitem{GuanLG12}
Hu~Guan, Huakang Li, and Minyi Guo.
\newblock Semi-sparse algorithm based on multi-layer optimization for
  recommendation system.
\newblock In {\em PMAM}, pages 148--155, 2012.

\bibitem{cfsurvey}
Mohd~Abdul Hameed, Omar~Al Jadaan, and S.~Ramachandram.
\newblock Collaborative filtering based recommendation system: A survey.
\newblock {\em International Journal on Computer Science and Engineering},
  4(5):859--876, 2012.

\bibitem{Han2004203}
Peng Han, Bo~Xie, Fan Yang, and Ruimin Shen.
\newblock A scalable p2p recommender system based on distributed collaborative
  filtering.
\newblock {\em Expert Systems with Applications}, 27(2):203 -- 210, 2004.

\bibitem{Han2004}
Peng Han, Bo~Xie, Fan Yang, Jiajun Wang, and Ruimin Shen.
\newblock A novel distributed collaborative filtering algorithm and its
  implementation on p2p overlay network.
\newblock In Honghua Dai, Ramakrishnan Srikant, and Chengqi Zhang, editors,
  {\em Advances in Knowledge Discovery and Data Mining}, volume 3056 of {\em
  Lecture Notes in Computer Science}, pages 106--115. Springer Berlin
  Heidelberg, 2004.

\bibitem{6064611}
T.J. Hansen, M.~Morup, and L.K. Hansen.
\newblock Non-parametric co-clustering of large scale sparse bipartite networks
  on the gpu.
\newblock In {\em Machine Learning for Signal Processing (MLSP), 2011 IEEE
  International Workshop on}, pages 1 --6, sept. 2011.

\bibitem{Harth}
Andreas Harth, Michael Bauer, and Bernd Breutmann.
\newblock Collaborative filtering in a distributed environment: an agent-based
  approach.
\newblock In {\em Technical Report. University of Applied Sciences
  W$\ddot{u}$ürzburg, Germany}, June 2001.

\bibitem{metrics}
Terveen~G. Herlocker~J., Konstan~J. and Riedl J.
\newblock Evaluating collaborative filtering recommender systems.
\newblock {\em ACM Transactions on Information Systems.}, Vol. 22(1):5--53,
  January 2004.

\bibitem{Hofmann_2004}
Thomas Hofmann.
\newblock Latent semantic models for collaborative filtering.
\newblock {\em ACM Transactions on Information Systems}, 22(1):89—115, 2004.

\bibitem{Holmes}
M.~Holmes, A.~Gray, and CL. Isbell.
\newblock Quic-svd: Fast svd using cosine trees.
\newblock In {\em In proc.of NIPS}, pages 673--680, 2008.

\bibitem{Crossdomainsurvey}
Marius~Kaminskas Ignacio Fern\'{a}ndez-Tobías, Iv\'{a}n~Cantador and Francesco
  Ricci.
\newblock Cross-domain recommender systems: A survey of the state of the art.
\newblock {\em Second Spanish Conference on Information Retrieval (CERI 2012)},
  2012.

\bibitem{imdb}
Internet Movie~Database (IMDb).
\newblock \url{http://www.imdb.com}.

\bibitem{Isaacman:2011:DRP:2043932.2043948}
Sibren Isaacman, Stratis Ioannidis, Augustin Chaintreau, and Margaret
  Martonosi.
\newblock Distributed rating prediction in user generated content streams.
\newblock In {\em Proceedings of the fifth ACM conference on Recommender
  systems}, RecSys '11, pages 69--76, New York, NY, USA, 2011. ACM.

\bibitem{jannachRS}
Dietmar Jannach, Markus Zanker, Alexander Felfernig, and Gerhard Friedrich.
\newblock {\em Recommender Systems An Introduction}.
\newblock Cambridge University Press, 2011.

\bibitem{CF_Hadoop}
Guangquan~Zhang Jing~Jiang, Jie~Lu and Guodong Long.
\newblock Scaling-up item-based collaborative filtering recommendation
  algorithm based on hadoop.
\newblock {\em World Congress Services (SERVICES), 2011 IEEE}, pages 490 --497,
  july 2011.

\bibitem{ClassRSsurvey}
V.~G.~Talwar K.~N.~Rao.
\newblock Application domain and functional classification of recommender
  systems - a survey.
\newblock {\em DESIDOC Journal of Library and Information Technology},
  28(3):17--35, 2008.

\bibitem{Kanagal:2012:SRS:2336664.2336669}
Bhargav Kanagal, Amr Ahmed, Sandeep Pandey, Vanja Josifovski, Jeff Yuan, and
  Lluis Garcia-Pueyo.
\newblock Supercharging recommender systems using taxonomies for learning user
  purchase behavior.
\newblock {\em Proc. VLDB Endow.}, 5(10):956--967, June 2012.

\bibitem{karydi_margaritis_1}
Efthalia Karydi and Konstantinos Margaritis.
\newblock Multithreaded implementation of the slope one algorithm for
  collaborative filtering.
\newblock {\em in Proceedings of 8th International conference on Artificial
  Intelligence Applications and Innovations, AIAI2012}, 2012.

\bibitem{karydi_margaritis_2}
Efthalia Karydi and Konstantinos Margaritis.
\newblock Parallel implementation of the slope one algorithm for collaborative
  filtering.
\newblock {\em in Proceedings of 16th Panhellenic Conference of Informatics,
  pci2012}, 2012.

\bibitem{Kato:2010:SKN:1844765.1845125}
Kimikazu Kato and Tikara Hosino.
\newblock Solving k-nearest neighbor problem on multiple graphics processors.
\newblock In {\em Proceedings of the 2010 10th IEEE/ACM International
  Conference on Cluster, Cloud and Grid Computing}, CCGRID '10, pages 769--773,
  Washington, DC, USA, 2010. IEEE Computer Society.

\bibitem{katosvdgpu}
K.Kato and T.Hosino.
\newblock Singular value decomposition for collaborative filtering on a gpu.
\newblock In {\em IOP Conference Series: Materials Science and Engineering 10
  012017}, 2010.

\bibitem{Kumar10collaborativeweb}
A.~Kumar and P.Thambidurai.
\newblock Collaborative web recommendation systems -- a survey approach,
  January 2010.

\bibitem{kumar}
Neeraj Kumar, Naveen Chilamkurti, and Jong-Hyouk Lee.
\newblock Distributed context aware collaborative filtering approach for p2p
  service selection and recovery in wireless mesh networks.
\newblock In {\em Peer-to-Peer Networking and Applications}, volume~5, Boston,
  December 2012. Springer US.

\bibitem{kwon}
Bongjune Kwon and Hyuk Cho.
\newblock Scalable co-clustering algorithms.
\newblock In Ching-Hsien Hsu, LaurenceT. Yang, JongHyuk Park, and Sang-Soo Yeo,
  editors, {\em Algorithms and Architectures for Parallel Processing}, volume
  6081 of {\em Lecture Notes in Computer Science}, pages 32--43. Springer
  Berlin Heidelberg, 2010.

\bibitem{5161058}
S.~Lahabar and P.J. Narayanan.
\newblock Singular value decomposition on gpu using cuda.
\newblock In {\em Parallel Distributed Processing, 2009. IPDPS 2009. IEEE
  International Symposium on}, pages 1 --10, may 2009.

\bibitem{Li:2011:SNT:1998076.1998131}
Ruifeng Li, Yin Zhang, Haihan Yu, Xiaojun Wang, Jiangqin Wu, and Baogang Wei.
\newblock A social network-aware top-n recommender system using gpu.
\newblock In {\em Proceedings of the 11th annual international ACM/IEEE joint
  conference on Digital libraries}, JCDL '11, pages 287--296, New York, NY,
  USA, 2011. ACM.

\bibitem{5693295}
Huizhi Liang, J.~Hogan, and Yue Xu.
\newblock Parallel user profiling based on folksonomy for large scaled
  recommender systems: An implimentation of cascading mapreduce.
\newblock In {\em Data Mining Workshops (ICDMW), 2010 IEEE International
  Conference on}, pages 154 --161, dec. 2010.

\bibitem{Link05distributedrecommender}
Hamilton Link, Jared Saia, Randall Laviolette, and Terran Lane.
\newblock Distributed recommender systems and the network topologies that love
  them, 2005.

\bibitem{Liu:2011:PPL:1961189.1961198}
Zhiyuan Liu, Yuzhou Zhang, Edward~Y. Chang, and Maosong Sun.
\newblock Plda+: Parallel latent dirichlet allocation with data placement and
  pipeline processing.
\newblock {\em ACM Trans. Intell. Syst. Technol.}, 2(3):26:1--26:18, May 2011.

\bibitem{LOU1}
G.~Louppe and P.~Geurts.
\newblock A zealous parallel gradient descent algorithm.
\newblock In {\em NIPS 2010 Workshop on Learning on Cores, Clusters and
  Clouds}, 2010.

\bibitem{Lü20121}
Linyuan Lü, Matúš Medo, Chi~Ho Yeung, Yi-Cheng Zhang, Zi-Ke Zhang, and Tao
  Zhou.
\newblock Recommender systems.
\newblock {\em Physics Reports}, 519(1):1 -- 49, 2012.
\newblock Recommender Systems.

\bibitem{metacritic}
Metacritic.
\newblock \url{http:// www.metacritic.com}.

\bibitem{MillerKR04}
Bradley~N. Miller, Joseph~A. Konstan, and John Riedl.
\newblock Pocketlens: Toward a personal recommender system.
\newblock {\em ACM Trans. Inf. Syst.}, 22(3):437--476, 2004.

\bibitem{RottenTomatoes}
Rotten~Tomatoes movie~review system.
\newblock \url{http://www.rottentomatoes.com}.

\bibitem{pcf_streaming_data}
Christopher C.~Johnson Muqeet~Ali and Alex~K. Tang.
\newblock Parallel collaborative filtering for streaming data.
\newblock {\em http://www.cs.utexas.edu/~cjohnson/}, December 2011.

\bibitem{IBM_India_10}
A.~Narang, R.~Gupta, A.~Joshi, and V.K. Garg.
\newblock Highly scalable parallel collaborative filtering algorithm.
\newblock {\em High Performance Computing (HiPC), 2010 International Conference
  on}, pages 1 --10, dec. 2010.

\bibitem{6413871}
Ankur Narang, Abhinav Srivastava, and Naga Praveen~Kumar Katta.
\newblock High performance offline and online distributed collaborative
  filtering.
\newblock In {\em Data Mining (ICDM), 2012 IEEE 12th International Conference
  on}, pages 549 --558, dec. 2012.

\bibitem{IBM_India_11}
Srivastava~Abhinav Narang~Ankur and Naga Praveen~Kumar Katta.
\newblock Distributed scalable collaborative filtering algorithm.
\newblock {\em Euro-Par'11 Proceedings of the 17th international conference on
  Parallel processing}, 2011.

\bibitem{Olsson98decentralisedsocial}
Tomas Olsson.
\newblock Decentralised social filtering based on trust.
\newblock In {\em In Working Notes of the AAAI-98 Recommender Systems
  Workshop}, pages 84--88. AAAI Press, 1998.

\bibitem{Park:2012:LRC:2181339.2181690}
Deuk~Hee Park, Hyea~Kyeong Kim, Il~Young Choi, and Jae~Kyeong Kim.
\newblock A literature review and classification of recommender systems
  research.
\newblock {\em Expert Syst. Appl.}, 39(11):10059--10072, September 2012.

\bibitem{Netflix_Prize}
Netflix Prize.
\newblock \url{http://www.netflixprize.com/}.

\bibitem{recht}
Benjamin Recht and Christopher Re.
\newblock Parallel stochastic gradient algorithms for large-scale matrix
  completion. submitted for publication.
\newblock 2011.

\bibitem{RechtRWN11}
Benjamin Recht, Christopher Re, Stephen~J. Wright, and Feng Niu.
\newblock Hogwild: A lock-free approach to parallelizing stochastic gradient
  descent.
\newblock In {\em NIPS}, pages 693--701, 2011.

\bibitem{Movielensds}
GroupLens Research.
\newblock Movielens data sets.
\newblock \url{http://www.grouplens.org/node/73}.

\bibitem{Resnick_Iacovou_Suchak_Bergstrom_Riedl_1994}
Paul Resnick, Neophytos Iacovou, Mitesh Suchak, Peter Bergstrom, and John
  Riedl.
\newblock {\em GroupLens : An Open Architecture for Collaborative Filtering of
  Netnews}, page 175—186.
\newblock ACM, 1994.

\bibitem{RicciHand}
Francesco Ricci, Lior Rokach, Bracha Shapira, and Paul~B. Kantor, editors.
\newblock {\em Recommender Systems Handbook}.
\newblock Springer, 2011.

\bibitem{cha_rs_riedl}
John Riedl.
\newblock Research challenges in recommender systems.
\newblock {\em Tutorial sessions Recommender Systems Conference ACM RecSys},
  October 2009.

\bibitem{Ripeanu02mappingthe}
Matei Ripeanu, Ian Foster, and Adriana Iamnitchi.
\newblock Mapping the gnutella network: Properties of large-scale peer-to-peer
  systems and implications for system design.
\newblock {\em IEEE Internet Computing Journal}, 6, 2002.

\bibitem{Ruffo:2009:PRS:1462159.1462163}
Giancarlo Ruffo and Rossano Schifanella.
\newblock A peer-to-peer recommender system based on spontaneous affinities.
\newblock {\em ACM Trans. Internet Technol.}, 9(1):4:1--4:34, February 2009.

\bibitem{atisha}
Atisha Sachan and Vineet Richariya.
\newblock A survey on recommender systems based on collaborative filtering
  technique.
\newblock {\em International Journal of Innovations in Engineering and
  Technology (IJIET)}, 2(2), 2013.

\bibitem{Salakhutdinov:2007:RBM:1273496.1273596}
Ruslan Salakhutdinov, Andriy Mnih, and Geoffrey Hinton.
\newblock Restricted boltzmann machines for collaborative filtering.
\newblock In {\em Proceedings of the 24th International Conference on Machine
  Learning}, ICML '07, pages 791--798, New York, NY, USA, 2007. ACM.

\bibitem{Sarwar_Karypis_Konstan_Riedl_2000}
Badrul~M Sarwar, George Karypis, Joseph~A Konstan, and John~T Riedl.
\newblock Application of dimensionality reduction in recommender system -- a
  case study.
\newblock {\em Architecture}, 1625(1):264—8, 2000.

\bibitem{sarwar_e_com}
Konstan~J. Sarwar~B., Karypis~G. and Riedl J.
\newblock Analysis of recommendation algorithms for e-commerce.
\newblock {\em ACM E-Commerce 2000 Conference}, October 2000.

\bibitem{Schelter:2012:SSN:2365952.2365984}
Sebastian Schelter, Christoph Boden, and Volker Markl.
\newblock Scalable similarity-based neighborhood methods with mapreduce.
\newblock In {\em Proceedings of the sixth ACM conference on Recommender
  systems}, RecSys '12, pages 163--170, New York, NY, USA, 2012. ACM.

\bibitem{Schelter:2013:DMF:2507157.2507195}
Sebastian Schelter, Christoph Boden, Martin Schenck, Alexander Alexandrov, and
  Volker Markl.
\newblock Distributed matrix factorization with mapreduce using a series of
  broadcast-joins.
\newblock In {\em Proceedings of the 7th ACM conference on Recommender
  systems}, RecSys '13, pages 281--284, New York, NY, USA, 2013. ACM.

\bibitem{survey_cft}
Xiaoyuan Su and Taghi~M. Khoshgoftaar.
\newblock A survey of collaborative filtering techniques.
\newblock {\em Advances in Artificial Intelligence}, Vol. 2009, January 2009.

\bibitem{Tang:2013:SMF:2487788.2487801}
Lei Tang and Patrick Harrington.
\newblock Scaling matrix factorization for recommendation with randomness.
\newblock In {\em Proceedings of the 22Nd International Conference on World
  Wide Web Companion}, WWW '13 Companion, pages 39--40, Republic and Canton of
  Geneva, Switzerland, 2013. International World Wide Web Conferences Steering
  Committee.

\bibitem{abs-1109-3318}
Dan-Cristian Tomozei and Laurent Massouli{\'e}.
\newblock Distributed user profiling via spectral methods.
\newblock {\em CoRR}, abs/1109.3318, 2011.

\bibitem{6337114}
A.~Tripathy, S.~Mohan, and R.~Mahapatra.
\newblock Optimizing a collaborative filtering recommender for many-core
  processors.
\newblock In {\em Semantic Computing (ICSC), 2012 IEEE Sixth International
  Conference on}, pages 261 --268, sept. 2012.

\bibitem{Tveit:2001:PBR:381461.381466}
Amund Tveit.
\newblock Peer-to-peer based recommendations for mobile commerce.
\newblock In {\em Proceedings of the 1st international workshop on Mobile
  commerce}, WMC '01, pages 26--29, New York, NY, USA, 2001. ACM.

\bibitem{10.1109/TLT.2012.11}
Katrien Verbert, Nikos Manouselis, Xavier Ochoa, Martin Wolpers, Hendrik
  Drachsler, Ivana Bosnic, and Erik Duval.
\newblock Context-aware recommender systems for learning: A survey and future
  challenges.
\newblock {\em IEEE Transactions on Learning Technologies}, 5(4):318--335,
  2012.

\bibitem{Bondhugula}
Bondhugula Vinay, Govindaraju Naga, and Manocha Dinesh.
\newblock Fast svd on graphics processors.
\newblock In {\em Tech. Rep., UNC Chapel Hill}, 2006.

\bibitem{pdp}
Emmanouil Vozalis and Konstantinos Margaritis.
\newblock Analysis of recommender systems' algorithms.
\newblock {\em The 6th Hellenic European Conference on Computer Mathematics \&
  its Applications (HERCMA), Athens, Greece}, pages 732--745, 2003.

\bibitem{Wang:2006:DCF:1141277.1141522}
Jun Wang, Johan Pouwelse, Reginald~L. Lagendijk, and Marcel J.~T. Reinders.
\newblock Distributed collaborative filtering for peer-to-peer file sharing
  systems.
\newblock In {\em Proceedings of the 2006 ACM symposium on Applied computing},
  SAC '06, pages 1026--1030, New York, NY, USA, 2006. ACM.

\bibitem{Wang:2009:PPL:1574036.1574062}
Yi~Wang, Hongjie Bai, Matt Stanton, Wen-Yen Chen, and Edward~Y. Chang.
\newblock Plda: Parallel latent dirichlet allocation for large-scale
  applications.
\newblock In {\em Proceedings of the 5th International Conference on
  Algorithmic Aspects in Information and Management}, AAIM '09, pages 301--314,
  Berlin, Heidelberg, 2009. Springer-Verlag.

\bibitem{weka}
Weka.
\newblock \url{http://www.cs.waikato.ac.nz/ml/weka/}.

\bibitem{multicore_cf}
Yao Wu, Qiang Yan, Danny Bickson, Yucheng Low, and Qing Yang.
\newblock Efficient multicore collaborative filtering.
\newblock {\em Matrix}, 2011.

\bibitem{Xie20071349}
Bo~Xie, Peng Han, Fan Yang, Rui-Min Shen, Hua-Jun Zeng, and Zheng Chen.
\newblock Dcfla: A distributed collaborative-filtering neighbor-locating
  algorithm.
\newblock {\em Information Sciences}, 177(6):1349 -- 1363, 2007.

\bibitem{mfrm}
Hsiang-Fu Yu, Cho-Jui Hsieh, Si~Si, and Inderjit~S. Dhillon.
\newblock Scalable coordinate descent approaches to parallel matrix
  factorization for recommender systems.
\newblock {\em In Proceedings of the IEEE International Conference on Data
  Mining (ICDM)}, pages 765--774, 2012.

\bibitem{Zastrau:2012:SGD:2406479.2406497}
David Zastrau and Stefan Edelkamp.
\newblock Stochastic gradient descent with gpgpu.
\newblock In {\em Proceedings of the 35th Annual German conference on Advances
  in Artificial Intelligence}, KI'12, pages 193--204, Berlin, Heidelberg, 2012.
  Springer-Verlag.

\bibitem{6384989}
Gao Zhanchun and Liang Yuying.
\newblock Improving the collaborative filtering recommender system by using
  gpu.
\newblock In {\em Cyber-Enabled Distributed Computing and Knowledge Discovery
  (CyberC), 2012 International Conference on}, pages 330 --333, oct. 2012.

\bibitem{Tagsurvey}
Zi-Ke Zhang, Tao Zhou, and Yi-Cheng Zhang.
\newblock Tag-aware recommender systems: A state-of-the-art survey.
\newblock {\em Journal of Computer Science and Technology}, 26:767--777, 2011.

\bibitem{Ub_Hadoop}
Zhi-Dan Zhao and Ming-Sheng Shang.
\newblock User-based collaborative-filtering recommendation algorithms on
  hadoop.
\newblock {\em 2010 Third International Conference on Knowledge Discovery and
  Data Mining}, pages 478 --481, jan. 2010.

\bibitem{Par_cf_netflix}
Yunhong Zhou, Dennis Wilkinson, Robert Schreiber, and Rong Pan.
\newblock Large-scale parallel collaborative filtering for the netflix prize.
\newblock {\em Algorithmic Aspects in Information and Management}, Vol.5034:337
  -- 348, 2008.

\bibitem{Zhuang:2013:FPS:2507157.2507164}
Yong Zhuang, Wei-Sheng Chin, Yu-Chin Juan, and Chih-Jen Lin.
\newblock A fast parallel sgd for matrix factorization in shared memory
  systems.
\newblock In {\em Proceedings of the 7th ACM conference on Recommender
  systems}, RecSys '13, pages 249--256, New York, NY, USA, 2013. ACM.

\bibitem{NIPS2010_1162}
Martin Zinkevich, Markus Weimer, Alex Smola, and Lihong Li.
\newblock Parallelized stochastic gradient descent.
\newblock In J.~Lafferty, C.~K.~I. Williams, J.~Shawe-Taylor, R.S. Zemel, and
  A.~Culotta, editors, {\em Advances in Neural Information Processing Systems
  23}, pages 2595 -- 2603, 2010.

\bibitem{NvidiaUM}
NVIDIA~Developer Zone.
\newblock Unified memory in cuda 6.
\newblock
  \url{https://devblogs.nvidia.com/parallelforall/unified-memory-in-cuda-6/}.

\end{thebibliography}
\bibliographystyle{plain}

\end{document}